\documentclass[jgrga,draft]{agutex2008}
\usepackage[pdftex]{graphicx}
\usepackage{color}
\usepackage{amssymb}
\usepackage{lineno}
\usepackage{gensymb}
\usepackage{amsmath}
\nolinenumbers
\authorrunninghead{ZHU ET AL.}
\def\titlerunninghead#1{\def\thetitle{{#1}}}
\titlerunninghead{AURORAL SIGNATURE OF BALLOONING INDUCED PLASMOID}
\journalid{}
\articleid{}{}
\paperid{}
\cpright{}{}
\received{} \revised{} \accepted{} \published{}
\authoraddr
{Ping Zhu, 1037 Luoyu Road, Wuhan, Hubei 430074, China; Jun Liang, 2500 University Drive, Calgary, Alberta T2N 1N4, Canada (zhup@hust.edu.cn; liangj@ucalgary.ca)\\}

\def\nms{\mathsurround=0pt}
\def\oversim#1#2{\lower 2pt\vbox{\baselineskip 0pt \lineskip 1pt
    \ialign{$\nms#1\hfil##\hfil$\crcr#2\crcr\sim\crcr}}}
\def\ltsim{\mathrel{\mathpalette\oversim<}} 
\def\gtsim{\mathrel{\mathpalette\oversim>}} 

\newcommand{\pop}{{\it Phys. Plasmas}}
\renewcommand{\jgr}{{\it J. Geophys. Res.}}

\renewcommand{\grl}{{\it Geophys. Res. Lett.}}

\newcommand{\procrsl}{{\it Proc. R. Soc. London}}

\newcommand{\jcp}{{\it J. Comput.Phys.}}
\newcommand{\ag}{{\it Ann. Geophys.}}

\newcommand{\jgrsp}{{\it J. Geophys. Res. Space Physics}}
\newcommand{\ess}{{\it Earth Space Sci.}}
\newcommand{\pss}{{\it Planet. Space Sci.}}
\newcommand{\fip}{{\it Front. Phys.}}
\newcommand{\apjss}{{\it Astrophys. J. Suppl. Ser.}}
\newcommand{\jastp}{{\it J. Atmos. Sol.-Terr. Phys.}}

\begin{document}
\title
{Auroral signatures of ballooning instability and plasmoid formation processes in the near-Earth magnetotail}

\authors{Ping Zhu \altaffilmark{1,2},
  Jun Liang \altaffilmark{3},
  Jiaxing Liu \altaffilmark{1},
  Sui Wan \altaffilmark{1}, and
  Eric Donovan \altaffilmark{3}
}

\altaffiltext{1}
{State Key Laboratory of Advanced Electromagnetic Technology,
International Joint Research Laboratory of Magnetic Confinement Fusion and
Plasma Physics, School of Electrical and Electronic Engineering,
Huazhong University of Science and Technology, Wuhan, 430074, China
}

\altaffiltext{2}
{Department of Nuclear Engineering and Engineering Physics, University of Wisconsin-Madison, Madison, Wisconsin 53706, USA}
                          
\altaffiltext{3}
{Department of Physics and Astronomy, University of Calgary, Calgary, Alberta T2N 1N4, Canada}

\begin{abstract}
The nonlinear development of ballooning instability and the subsequently induced plasmoid formation in the near-Earth magnetotail demonstrated in MHD simulations has been proposed as a potential trigger mechanism for substorm onset over the past decade, and their connections to the in-situ satellite and ground all-sky auroral optical observations have been a subject of continued research. In this work, a set of THEMIS substorm onset events with good conjunction of auroral observations has been selected for comparative simulation study, whose pre-onset magnetotail configuration and conditions are inferred from in-situ data and compared with the onset conditions of ballooning instability obtained in our MHD simulations. The evolution of the near-Earth magnetotail is followed, where the signatures of ballooning instability and the plasmoid formation are extracted from simulations and compared with the magnetic fields and flow patterns within the magnetotail region from observation data. The field-aligned current (FAC) density is evaluated at the Earth side boundary of the magnetotail domain of simulation, which is further mapped along magnetic field lines to the auroral ionosphere and compared with the auroral pattern and evolution there in terms of growth rate, dominant wavenumber, and absolute auroral intensities. Such validation efforts are also the first step towards the development of a self-consistent coupling model that includes the magnetotail-ionosphere interaction in the substorm onset process.
\end{abstract}

\begin{article}
\section{Introduction}
Ballooning instability is a pressure-gradient-driven mode in magnetized plasma localized in the unfavorable curvature regions of magnetic ﬁeld lines. It is a ubiquitous and fundamental process involved in many phenomena in natural and laboratory magnetized plasmas with high $\beta$ values. Here $\beta$ is the ratio of plasma and magnetic pressures. Having long been proposed as a potential candidate mechanism for the trigger of substorm onset, the ballooning instability of the Earth's magnetotail has received continued interest and investigation in theory and simulations over the past decades. Early ballooning analyses on magnetotail observations often directly adopt the concept and formula of interchange instability from tokamak plasma theory (e.g.~\citep{bernstein58a,sonnerup63a,chengaf85a,fazakerley93a,chencx93a} )  or the ballooning mode dispersion relation based on the local approximation with various non-ideal effects~\citep{miura89a,pu92a,ohtani93a,cheng98a}.  Later \citet{hameiri91a} extended the tokamak ballooning mode theory to the magnetotail plasma configuration, where special attentions are given to its field aligned eigenmode properties subject to the ionospheric boundary conditions.  Subsequent studies on the linear ballooning stability are mostly based on the non-local eigenmode analysis with the ionospheric boundary conditions taken into account. Notably, \citet{leedy92a} pointed out that the ﬂux tubes in the magnetotail are line-tied to the ionosphere, and that such a constraint is stabilizing due to the nonzero plasma compression and the enhanced line-binding force~\citep{hurricane95a,hurricane96a,hurricane97a,bhattacharjee98a,ma98a}.  Thus the ionosphere boundary condition plays a nontrivial role in determining the ballooning mode stability of the near-Earth magnetotail, which may be one of the most direct manifestations of the magnetotail-ionospheric coupling and interaction during the substorm onset process.

In fact, the original speculation on the ballooning instability of the Earth's magnetotail as a potential trigger mechanism for substorm onset was inspired by early auroral observations.  One such early observation of auroral structure that is associated with ballooning instability may be the westward traveling surge (WTS) captured by the all-sky camera~\citep{roux91a}. The field aligned current (FAC) is suggested as the mechanism connecting the WTS in auroral zone to the interchange type of instability developed in the transition zone between the dipole-like and the tail-like regions of magnetotail. \citet{elphinstone95a} provided perhaps the first evidence of azimuthal-spaced auroral structures along breakup arcs and hypothesized their possible linkage to ballooning-interchange instabilities, though their observations are limited to $1$-minute resolution. Taking advantage of the $3$-second resolution of THEMIS ASI, ~\citet{donovan06a,liang08a} made definitive identification of the azimuthal quasi-periodic auroral structures (a.k.a. “beading”) on a segment of pre-existing breakup arc shortly prior to the substorm expansion onset. Similar observations were also made by~\citet{henderson09a}, in which the beading and the subsequent poleward expansion are clearly captured in the IMAGE data during a substorm onset event. The magnetic conjugacy of northern and southern auroral bead structures was observed and established later~\citep{motoba12a,motoba12b,motoba15a}, which strongly suggests the presence of a common driver of these beads in the magnetotail equatorial region. The characteristic azimuthal wavelength and exponential growth rate of these onset beads have been thoroughly investigated from optical observations~\citep{liang08a,kalmoni15a,kalmoni17a,nishimuray16b}. Such temporal-spatial scales of these quasi-periodic auroral structures cast observational constraints on substorm onset theories.

By now, the auroral bead has become an established precursor for auroral substorm onset, and has been dubbed by many as alluding to magnetotail instabilities, though different versions of ballooning instabilities were invoked in the literature. After the presence of ballooning instability was found in the magnetotail prior to the substorm onset in the global simulation of a realistic substorm event using OpenGGCM~\citep{raeder10c}, the mapping projection of the ballooning mode structure from the equatorial plane to the ionosphere was able to reproduce certain qualitative features of the beading structure observed in the auroral zone~\citep{raeder12a}. \citet{kalmoni17a} compared a few candidate instability theories with their derived arc wave parameters, and concluded the shear flow ballooning instability \citep{voronkov97a} appears to best fit the observations. \citet{nishimuray16b} used the kinetic ballooning model developed by \citet{pritchett13a} to interpret their auroral beads/rays observations.  More recently, the projection and connection between the ballooning-interchange instability in the near-Earth plasma sheet and auroral beads have been further modeled and demonstrated in global magnetosphere simulations using the newly developed MHD code GAMERA~\citep{zhangb19a,sorathia20a}, a successor to the Lyon-Fedder-Mobarry (LFM) code~\citep{lyonj04a}, which are also compared with THEMIS ASI observations of the beading auroral structures with promising agreement. \citet{oberhagemann20a,babu24a} suggested that the development of an increasingly parallel anisotropic state during magnetotail stretching in the growth phase triggers a transition to ballooning instability, resulting in onset. We also mention that~\citet{lui16a} has argued that \citet{kalmoni15a}’s and~\citet{liang08a}’s results could also be interpreted via a cross-field current instability.

Whereas both observations and simulations have made solid progress towards establishing convincing connection between the auroral bead structure and the ballooning instability in the near-Earth magnetotail, how such a connection may further evolve and contribute to the subsequent substorm expansion has remained less clear. Recently, \citet{nishimuray25a} investigated the evolution of substorm onset beads into poleward expansion via THEMIS observations. They found that a thin arc appeared immediately poleward of the onset arc shortly after substorm onset but prior to significant poleward expansion. Poleward expansion occurred step-wise, with each step associated with a re-intensification of the poleward arc. \citet{nishimuray25a} suggested that the observations indicate a transition from near-Earth instability to the formation of a near-Earth neutral line (NENL), whose initial location was estimated to be ~11.8 RE. \citet{nishimuray25a}’s observations and proposals are conceptually similar to our previous MHD simulation results regarding the evolution of ballooning instability and the subsequent formation of NENL~\citep{zhu13d,zhu14a}.

In this work,  from the comparison between simulation and observation data in the auroral zone around the time of substorm onsets, we attempt to identify the auroral evolution that is associated with ballooning instability and its induced plasmoid formation processes in the near-Earth magnetotail. For such a purpose, we carry out meso-scale 3D MHD simulations using the NIMROD code~\citep{sovinec04a}. These simulations are able to recover and isolate the ballooning instability on closed field lines. We demonstrate that the three-dimensional (3D) processes such as the ballooning instability can significantly enhance the initiation of fast reconnection and induce the formation of plasmoids in the closed field line regions of near-Earth magnetotail. The simulations are set up according to the observed conditions of magnetotail during a realistic substorm onset event, and the field-aligned current from the simulation results are projected to the auroral zone to deduce the auroral structure evolution using the TREx-ATM (Auroral transport model) calculations~\citep{liang16a,liang17a,liang21a,liang25a}. The reconstructed auroral results are then compared with the THEMIS-ASI observation data. This effort can further reveal the magnetotail origin of the auroral activities following the initial appearance of beading structures and leading to the substorm expansion.

The rest of the paper is organized as follows. In Sec.~\ref{sec:themis}, an overview of the particular THEMIS substorm event we study is presented. In Sec.~\ref{sec:tail} , we first introduce the MHD simulation model and setup for the magnetotail region and then present two simulation cases for the nonlinear ballooning processes.  The FACs induced from magnetotail simulations are further projected to the auroral zone to reconstruct the auroral structures using the TREx-ATM model in Sec.~\ref{sec:iono}, which are followed by comparisons with observation data. Finally, we give a summary and discussion on our main findings in Sec.~\ref{sec:sum}.

\section{THEMIS substorm event overview}
\label{sec:themis}

\begin{figure}
  \vspace{-1in}
  \begin{center}
    \includegraphics[width=0.45\textwidth]{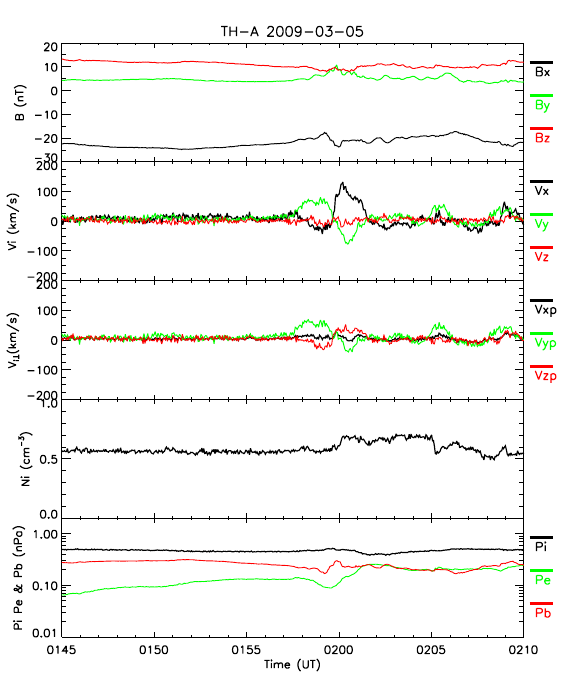}
    
    \includegraphics[width=0.45\textwidth]{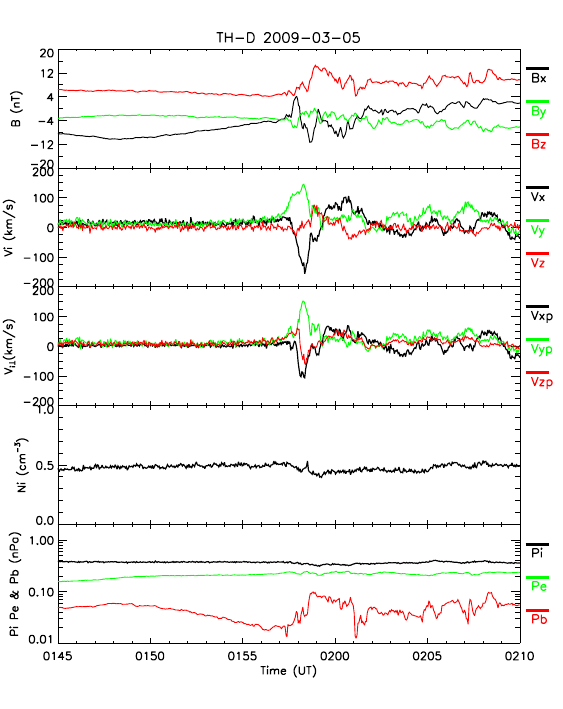}
    \includegraphics[width=0.45\textwidth]{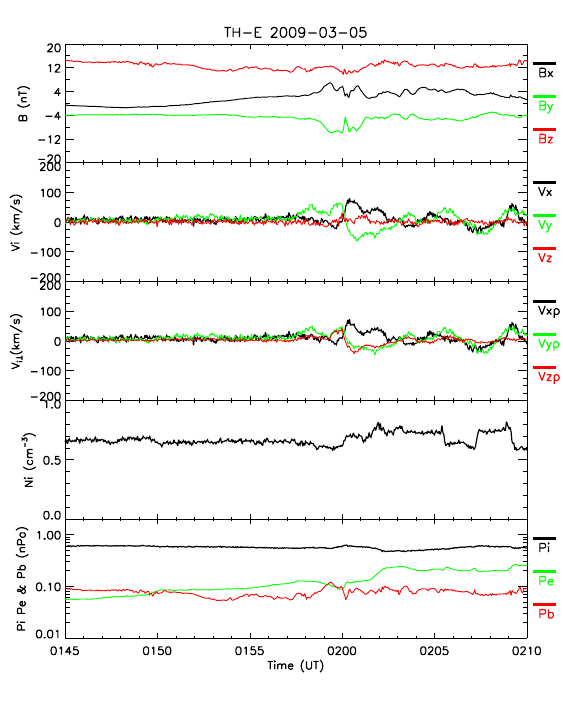}
  \end{center}
  \caption{THEMIS-A (a: upper), THEMIS-D (b: lower left) and THEMIS-E (c: lower right) observations of magnetic field (1st row),  ion velocity (2nd row), perpendicular ion velocity (3rd row), ion number density (4th row) and ion/magnetic pressure (5th row) as functions of time during the 2009, March 5 (``20090305'') substorm event.}
  \label{fig:themis_ade}
\end{figure}

\begin{figure}
  \begin{center}
      \includegraphics[width=0.49\textwidth]{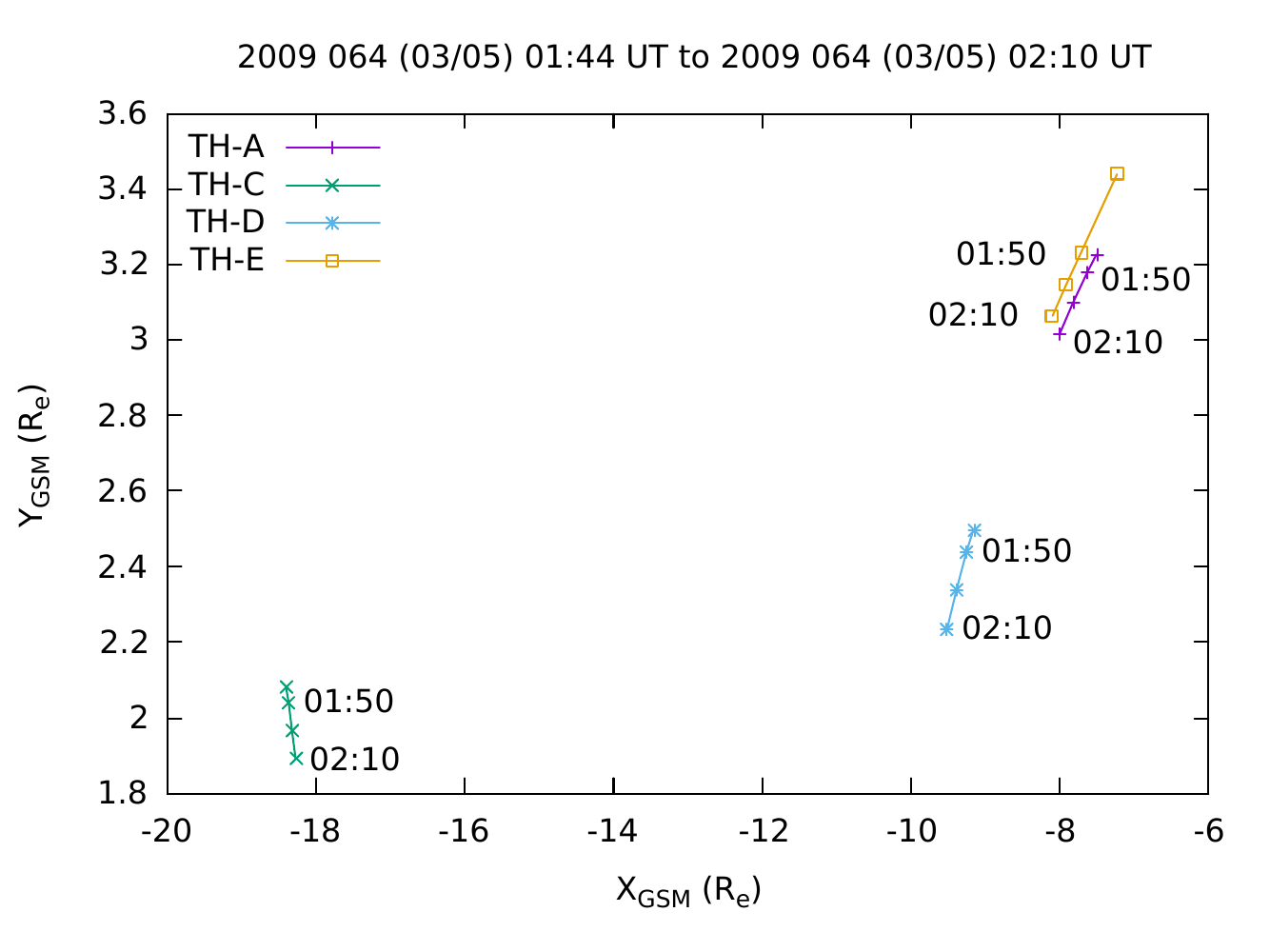}
      \includegraphics[width=0.49\textwidth]{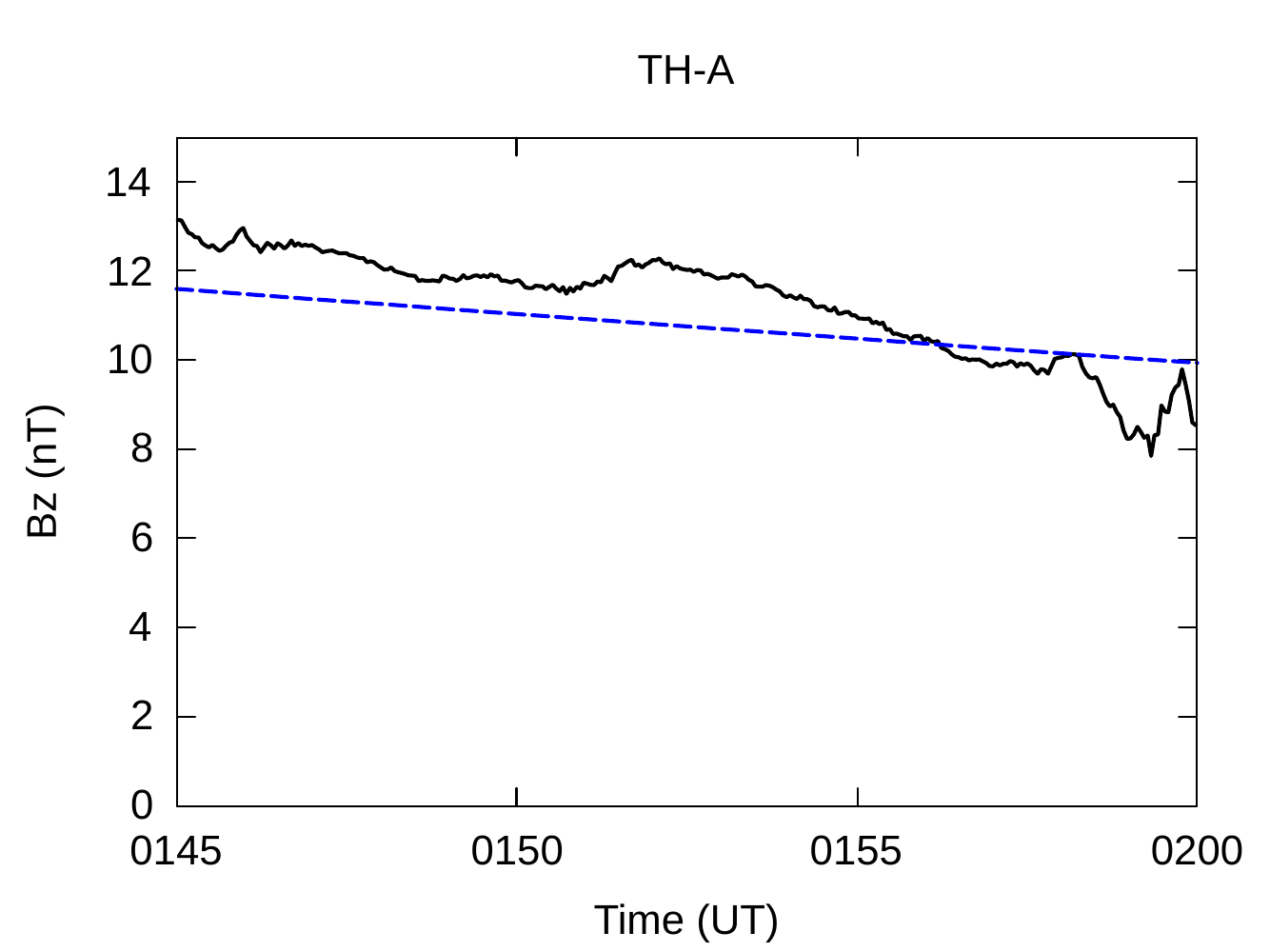}

      \includegraphics[width=0.49\textwidth]{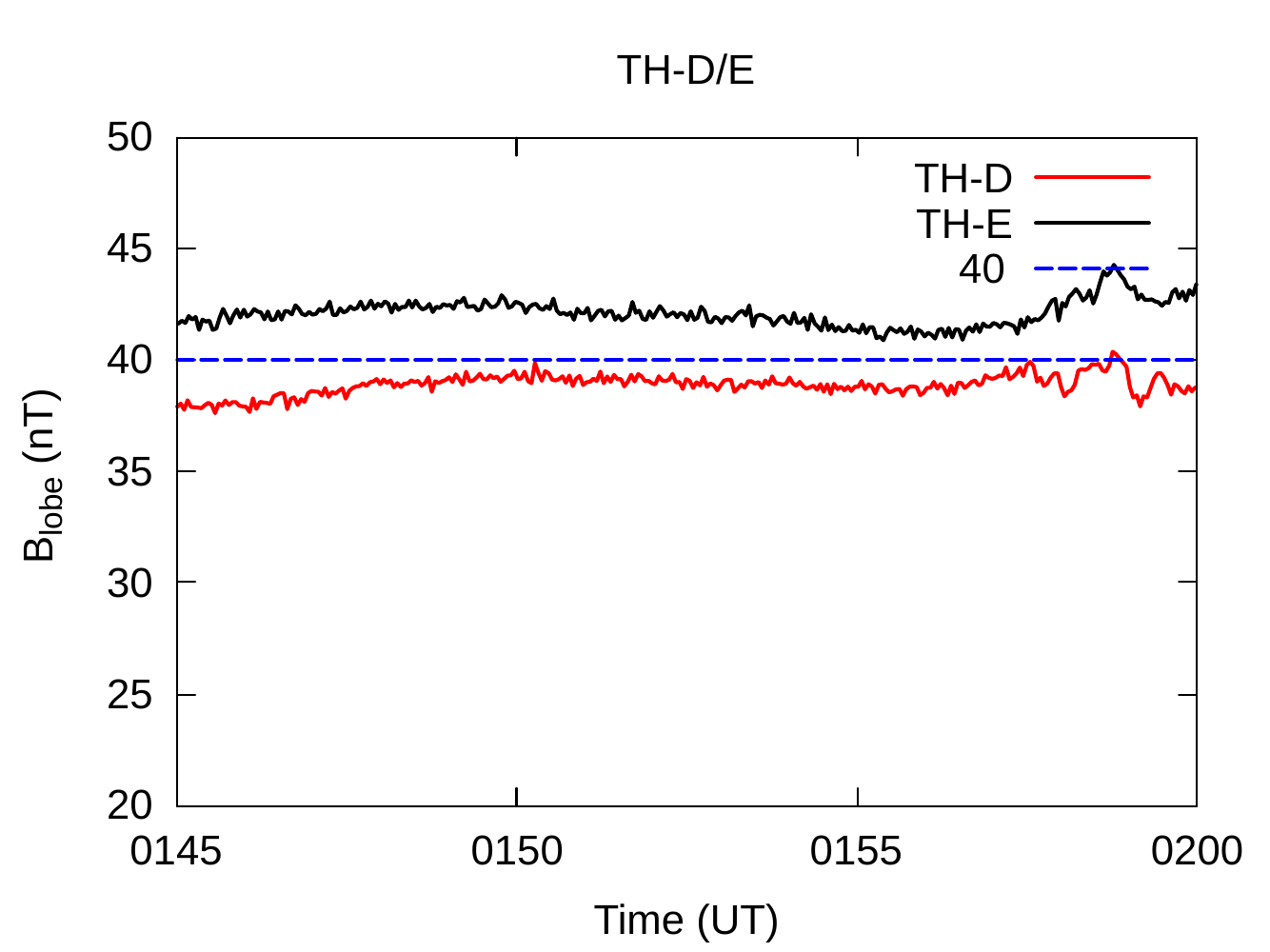}
      \includegraphics[width=0.45\textwidth]{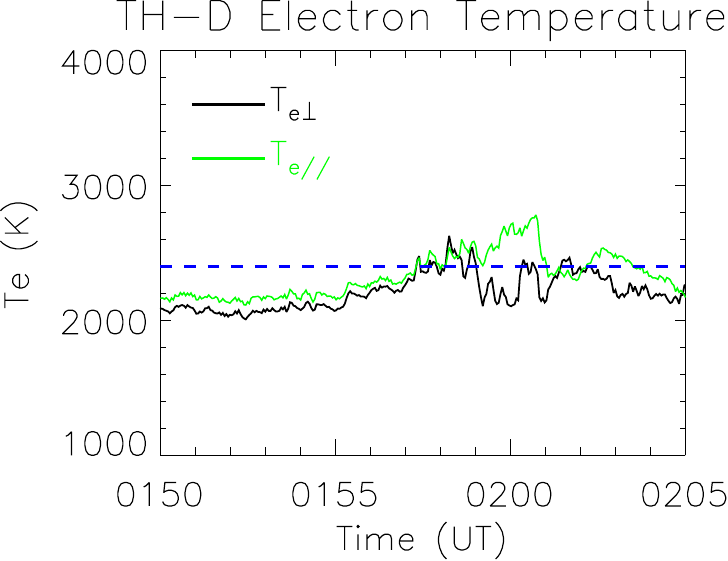}
  \end{center}
  \caption{(a: upper left) The ($X_{\rm GSM}$, $Y_{\rm GSM}$) coordinates of THEMIS satellite orbit trajectories; (b: upper right) the $B_z$ component of magnetic field measured by TH-A  (dark solid line) and the fitted value of $B_z$  (blue dashed line), (c: lower left) the lobe magnetic field magnitude $B_{\rm lobe}$ measured by TH-D (red solid line) and TH-E (dark solid line) and the fitted value of $B_{\rm lobe}=40nT$ (blue dashed line), and (d: lower right) the perpendicular (dark solid line) and parallel (green solid line) electron temperature measured by TH-D and the fitted electron temperature (blue dashed line) as functions of time during the 2009, March 5 (``20090305'') substorm event.}
  \label{fig:themis_orbit}
\end{figure}

\begin{figure}
\begin{center}
\includegraphics[width=\textwidth]{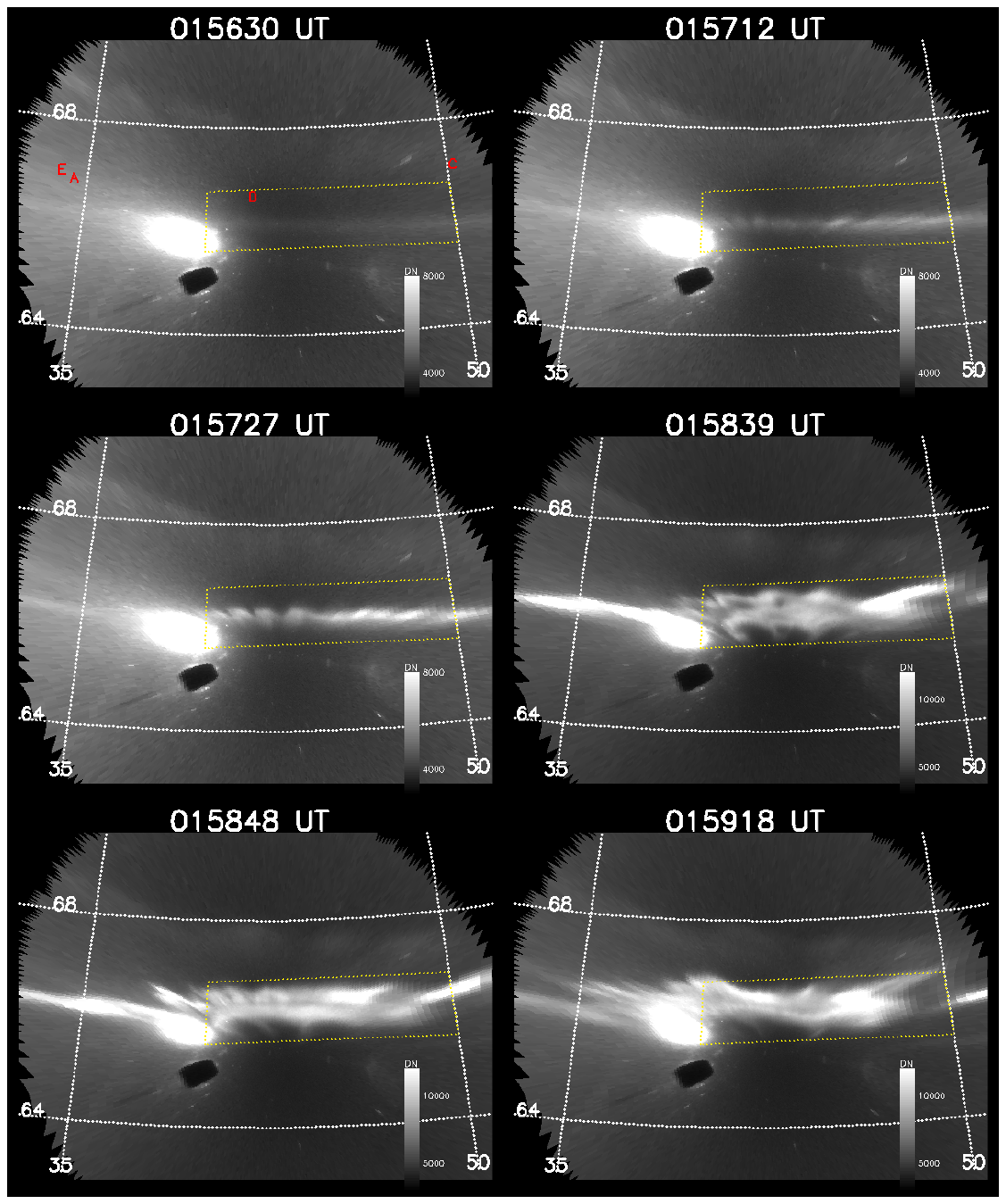}
\end{center}
\caption{ASI observation of auroral emission from 015630 UT to 015918 UT during the 20090305 substorm event .}
\label{fig:asi_20090305}
\end{figure}

We first present an overview of the THEMIS substorm events shown in Figs.~\ref{fig:themis_ade} and~\ref{fig:themis_orbit}. The substorm event under investigation occurred on March 5, 2009 at $\sim 2$ UT. It is a small substorm event occurring during an otherwise fairly quiet ($K_p=2-$) interval.  The substorm auroral breakup was captured by THEMIS ASI at KUUJ, though the eastern portion of the ASI field-of-view was partly contaminated by moonlight. The GSM location of four THEMIS spacecraft, A, B, D, and E, are shown in Fig.~\ref{fig:themis_orbit}a. The four spacecraft  were situated within or close to the onset sector (see Fig.~\ref {fig:asi_20090305}). In particular, THEMIS-D mapped fairly close to the breakup arc and the beading region of interest. Here, the observations of THEMIS-D/E are used to determine a few key parameters for the equilibrium current sheet configuration adopted in our subsequent simulation. Specifically, Fig.~\ref{fig:themis_orbit}b shows the $B_z$ field observed by the innermost spacecraft THEMIS-A. In GSM coordinates, the spacecraft moves gradually from $X_{\rm GSM}=-7.54 R_{\rm E}$ at  $0145$ UT to $X_{\rm GSM}=-7.83 R_{\rm E}$ at $0200$ UT, and the observed decreasing trend of $B_z$ is interpreted as spatial and is adopted to constrain the model equilibrium $B_z$ profile in Eq.~(\ref{eq:bzmin_pw}) in our simulation (shown as a blue dashed line ). Fig.~\ref{fig:themis_orbit}c shows the total pressure (the sum of ion pressure, electron pressure, and magnetic pressure) observed in THEMIS-D and THEMIS-E. These values are referenced to determine the lobe magnetic field strength ($40 nT$, shown as blue dashed line) of our model current sheet. The initial plasma density is also set according to the realistic THEMIS observations, i.e. $0.55 cm^{-3}$ at equator in the inner edge of current sheet close to THEMIS-A (see Fig.~\ref{fig:themis_ade}a).

Fig.~\ref{fig:asi_20090305} shows a few selected frames of the optical auroras for the substorm onset at  \(\sim \)2 UT on 5 March 2009, based on THEMIS white-light ASI data at the Kuujjuaq station. In this study, we will focus on the first few minutes of the onset. Typical of substorm breakup, beading structures emerge at \(\sim \)0157 UT along a pre-existing growth phase arc. Those beading structures intensify in the following minute and then begin to expand poleward. It is interesting to note that the poleward expansion does not proceed as a continuum. Instead, about ~1.5 min after the initial beading, an adjacent yet narrowly gaped new “arc” structure emerges \(\sim \)0.2\degree~poleward of the breakup arc. Just like the initial breakup arc, this new “arc” is not homogeneous but also features azimuthal structures and dynamic variations.

\section{MHD simulation of magnetotail ballooning instability}
\label{sec:tail}
The near-Earth magnetotail plasma is modeled using the resistive MHD equations implemented in the NIMROD code~\citep{sovinec04a} 
\begin{equation}
\frac{\partial\rho}{\partial t} + \nabla\cdot(\rho{\mathbf u})=0  \label{eq:den}
\end{equation}
\begin{equation}
\rho\left(\frac{\partial{\mathbf u}}{\partial t}+{\mathbf u}\cdot\nabla{\mathbf u}\right)
 = {\mathbf J}\times{\mathbf B} - \nabla p + \mu\nabla\cdot(\rho{\mathbf w}) \label{eq:vel}
\end{equation}
\begin{equation}
\frac{\partial p}{\partial t} + {\mathbf u}\cdot\nabla p
= -\gamma p\nabla\cdot{\mathbf u} \label{eq:tem}
\end{equation}
\begin{equation}
\frac{\partial{\mathbf B}}{\partial t}=-\nabla\times{\mathbf E} \label{eq:mag}
\end{equation}
\begin{equation}  
{\mathbf E} = -{\mathbf u}\times{\mathbf B} + \eta{\mathbf J} \label{eq:ohm}
\end{equation}
\begin{equation}  
\mu_0{\mathbf J}=\nabla\times{\mathbf B} \label{eq:cur}
\end{equation}
where $\rho$ is the mass density, ${\mathbf u}$ the plasma flow velocity, $p$ the pressure, ${\mathbf E}$ the electric field, ${\mathbf B}$ the magnetic field, ${\mathbf J}$ the current density, the adiabatic index or specific ratio $\gamma=5/3$, and ${\mathbf w}=\nabla{\mathbf u}+(\nabla{\mathbf u})^{\rm T}-\frac{2}{3}{\mathbf I}\nabla\cdot{\mathbf u}$ is the rate-of-strain tensor. In a weakly collisional or collisionless plasma the effective resistivity $\eta$ {and viscosity $\mu$} are small in absence of anomalous sources. The above set of equations has been implemented in both the linear and the fully nonlinear version in the NIMROD code~\citep{sovinec04a} used in our computation. A solid, no-slip wall boundary condition has been imposed on the sides of the computation domain in both $X_{\rm GSM}$ and $Z_{\rm GSM}$ directions, so that any potential influence from an external driver or inward flow may be excluded. The boundary condition in the $Y_{\rm GSM}$ direction is periodic. The spatial and temporal variables are normalized with the equilibrium scale length (e.g. Earth radius) and the Alfv\'{e}nic time $\tau_{\rm A}$, respectively.

\subsection{Initial equilibrium and perturbations}
\label{sec:initial}
\begin{figure}
\begin{center}
  \includegraphics[width=0.7\textwidth]{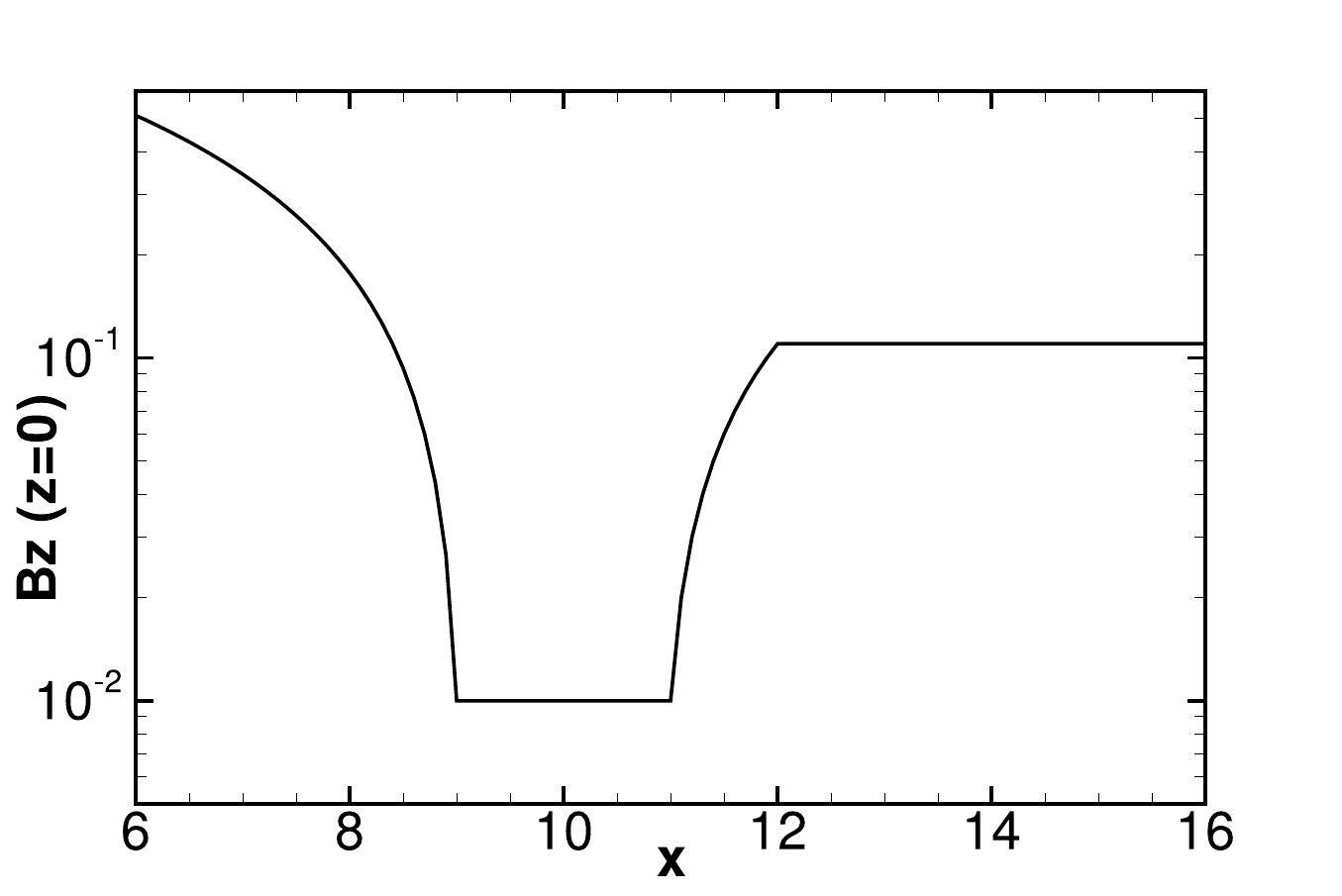}

  \includegraphics[width=0.7\textwidth]{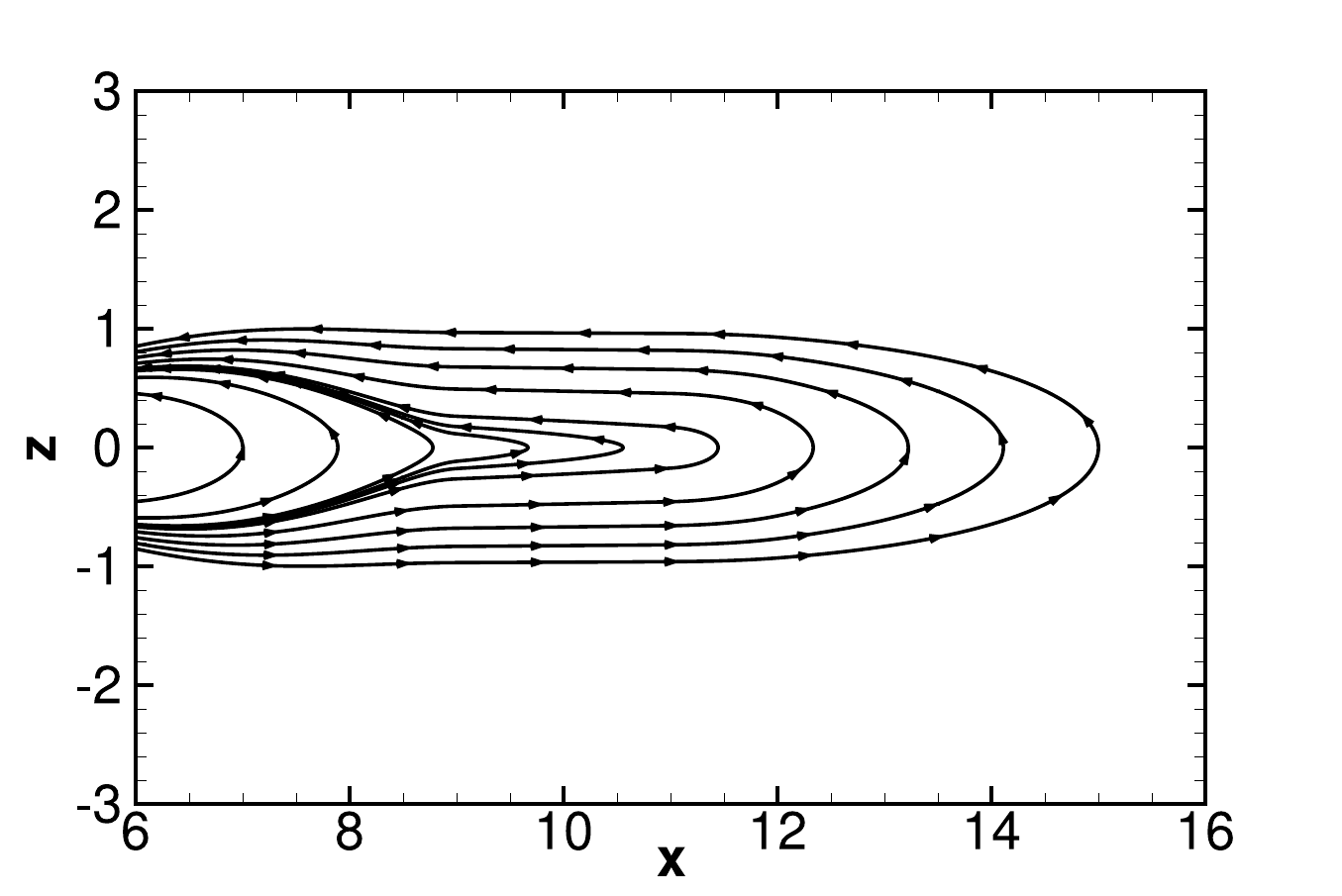}
\end{center}
\caption{Equatorial $B_z(z=0)$ profile as a function of $x$ (upper) and magnetic field streamlines in the meridian plane ($y=0$) (lower).}
\label{fig:gharris}
\end{figure}

To focus on the dynamics of ballooning instability and the induced plasmoid formation, we adopt the generalized Harris (GH) sheet as the model for the initial configuration of the near-Earth magnetotail ($\sim 6-30 R_{\mathrm E}$) during the slow growth phase of substorm~\citep{zhu13d,zhu14a}. In the normalized coordinates $(x,y,z)=(-X_{\rm GSM},-Y_{\rm GSM},Z_{\rm GSM})/R_{\rm E}$, the GH current sheet is specified as
\begin{equation}
{\mathbf B}_0(x,z)={\mathbf e}_y\times\nabla\Psi(x,z),
\end{equation}
\begin{equation}
\Psi(x,z)=-\lambda\ln{\frac{\cosh{\left[\frac{F(x)z}{\lambda}\right]}}{F(x)}},
\end{equation}
\begin{equation}
\ln{F(x)}=-\int \frac{B_{0z}(x,0)}{\lambda}dx,
\end{equation}
where $\lambda$ is the current sheet width, ${\mathbf e}_y$ the unit vector in $y$ direction, and all other symbols are conventional. Thus a generalized Harris sheet is to a large extent determined by the profile of $B_n=B_{0z}(x)$, where $B_n$ is the magnetic component normal to neutral sheet or equatorial plane at $z=0$. For the simulation of ballooning unstable magnetotail configuration, we  adopt a piece-wise linear function of $x$ for $B_{0z}(x,0)$ as~\citep{zhu13d,zhu14a}
\begin{equation}
B_{0z}(x,0)=\left\{
\begin{array}{ll}
\epsilon_1\frac{x-x_1}{x_1-x_0}+B_{\rm min} & x_0<x<x_1 \\
B_{\rm min} & x_1<x<x_2 \\
\epsilon_2\frac{x-x_2}{x_3-x_2}+B_{\rm min} & x_2<x<x_3 \\
\epsilon_2+B_{\rm min} & x>x_3
\end{array}
\right.,
\label{eq:bzmin_pw}
\end{equation}
where $x_i$ $(i=0,1,2,3)$, $\epsilon_i$ $(i=1,2)$, and $B_{\rm min}$ are parameters that define the profile of $B_{0z}(x,0)$. A schematic illustration of the $B_z$ profile and the corresponding magnetic field lines for the above magnetotail configuration is shown in Fig.~\ref{fig:gharris} for example.

The initial perturbation is composed of single or multiple Fourier components in the $y$ direction, where the wavenumber $k_y=2n\pi/L_y$ for each Fourier component, $L_y$ is the domain size in $y$, and $n$ is a non-zero integer.  To follow the evolution of ballooning instability of the above initial magnetotail configuration, a bi-quintic rectangular finite element mesh of $64\times 64$ with a polynomial degree of $5$ in each direction is used for the $x-z$ domain of $x\in [6,26],z\in [-3,3]$. In the $y$ direction, $64$ Fourier collocation points are used to resolve Fourier components in the range of $0\le k_yL_y/2\pi\le 20$.  The computation domain is periodic in the $y$ direction at boundary, whereas the solid no-slip boundary conditions are imposed on the other sides of the domain in both $x$ and $z$ directions, so that any potential influence from an external driver or inward flow may be excluded. 

\subsection{Magnetotail ballooning  instability evolution}
\label{sec:simruns}
In this subsection, we first present the NIMROD simulation results using the MHD equations~(\ref{eq:den})-(\ref{eq:cur}) on the magnetotail ballooning instability and plasmoid formation, starting from the initial equilibrium and perturbations introduced in the previous section. We then derive the corresponding evolution of field-aligned current (FAC) from the tail simulation results, whose auroral zone projection is input to TREx model for calculating the associated auroral signals later in Sec.~\ref{sec:fac_proj}.

\subsubsection{Single-mode initial perturbation}
\label{sec:single}
\begin{figure}
\begin{center}
\includegraphics[width=\textwidth]{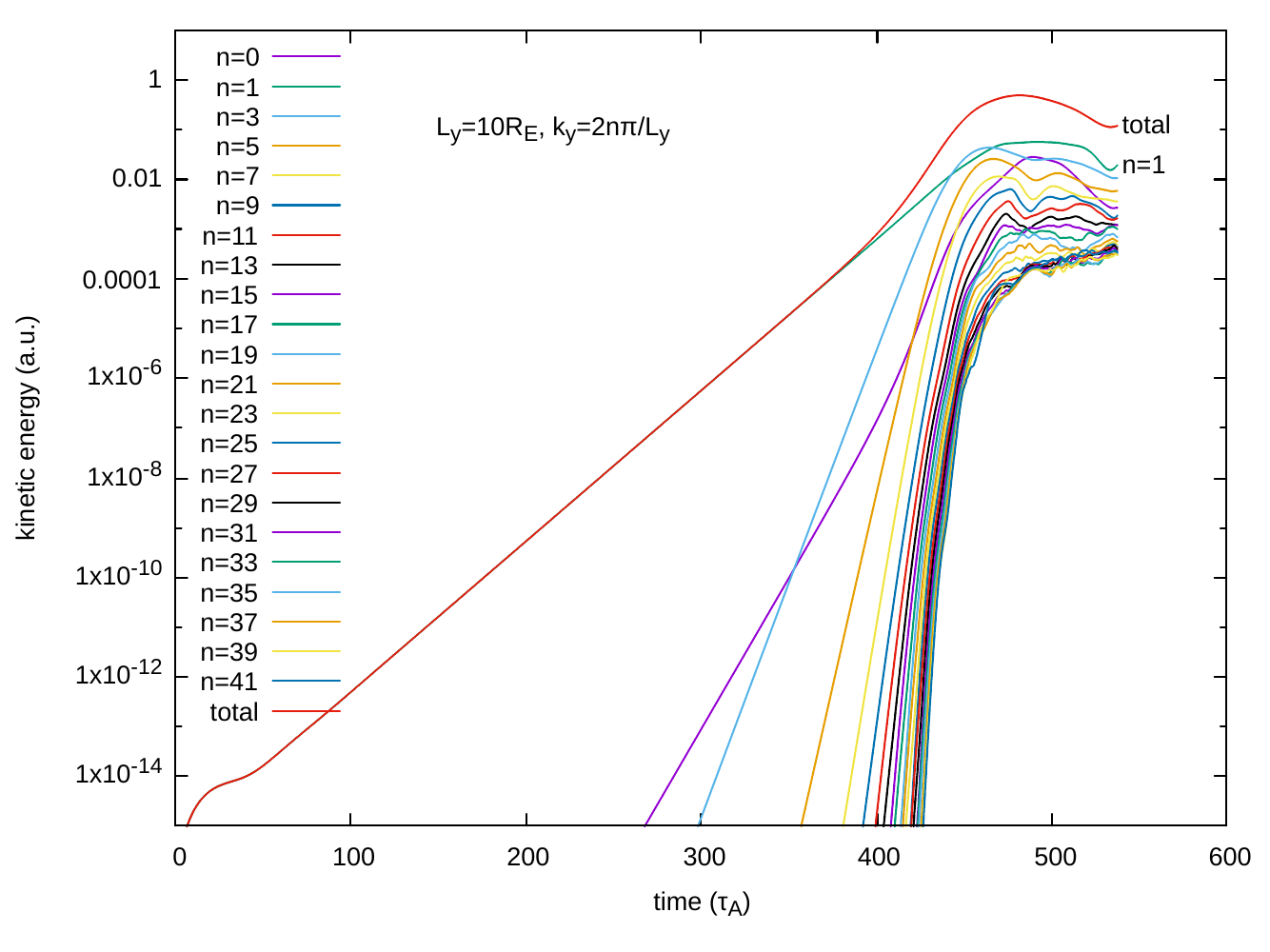}
\end{center}
\caption{Kinetic energies of  Fourier components of perturbation as well as their sum as functions of time, where only $n=1$ component has non-zero amplitude of $4.5\times 10^{-9}T$ in the initial perturbation.}
\label{fig:sm_ken_growth}
\end{figure}

We first consider the scenario where a monochromatic perturbation with a single Fourier component initiates in the near-Earth magnetotail. In particular, the initial magnetic perturbation has an amplitude of $4.5\times 10^{-9}T$ and a wavenumber $k_y=2\pi n/L_y$ where $n=1$ and $L_y=10R_e$.  The perturbation assumes a Gaussian profile along the $x$-axis, which is peaked around $x=-10R_e$ with half-height width of $10R_e$ as well. Although only the fundamental harmonic perturbation (i.e. $n=1$) is initialized with a nonzero amplitude, the nonlinear simulation includes and resolves a total of $43$ Fourier components with $n=0-42$ in the $y$-direction.  By the time the fundamental $n=1$ component enters  the middle of its initial exponential growth phase,  other Fourier components are sequentially excited and grow exponentially through nonlinear coupling and beating (Fig.~\ref{fig:sm_ken_growth}). Eventually all Fourier components reach saturation after about $450\tau_A$, when the amplitude of the fundamental $n=1$ mode remains dominant even though those of the $n\gtsim2$ components have become comparable. 

\begin{figure}
\begin{center}
  \includegraphics[width=0.85\textwidth]{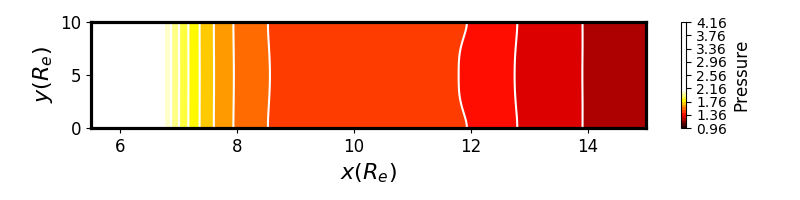}
  
  \includegraphics[width=0.85\textwidth]{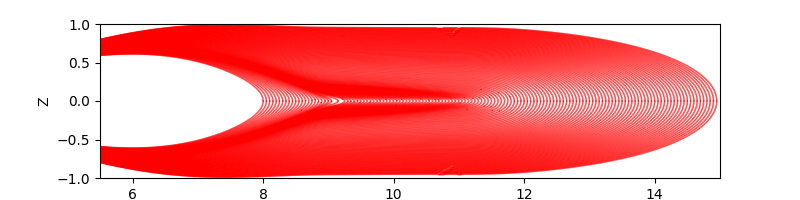}

  \includegraphics[width=0.85\textwidth]{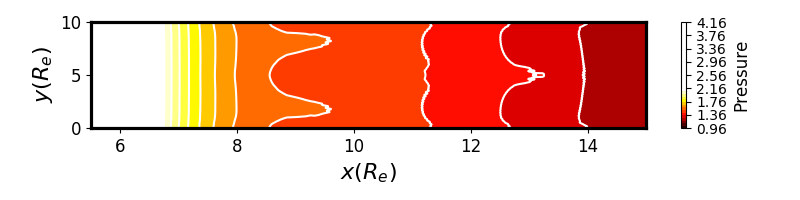}
  
  \includegraphics[width=0.85\textwidth]{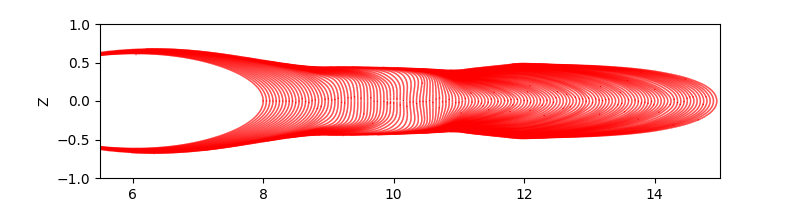}
  
  \includegraphics[width=0.85\textwidth]{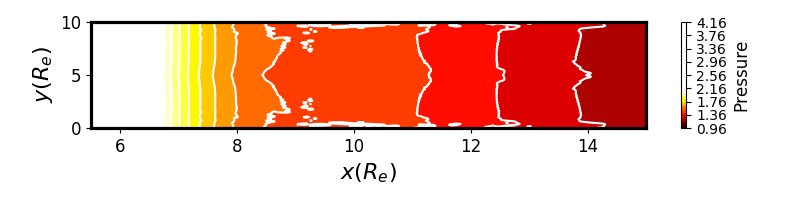}
  
  \includegraphics[width=0.85\textwidth]{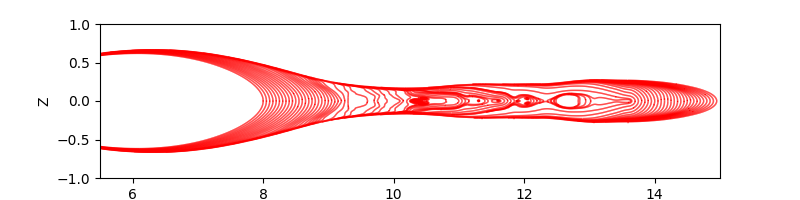} 
\end{center}
\caption{Pressure contours in $x-y$ equatorial plane ($z=0$) (1st, 3rd, and 5th panels) and magnetic field lines in $x-z$ meridian plane ($y=0$) (2nd, 4th, and 6th panels) at times $t=400$ (1st and 2nd panels), $t=470$ (3rd and 4th panels), and $t=504$ (5th and 6th panels).}
\label{fig:sm_pres_bline}
\end{figure}

\begin{figure}
\begin{center}
  \includegraphics[width=\textwidth]{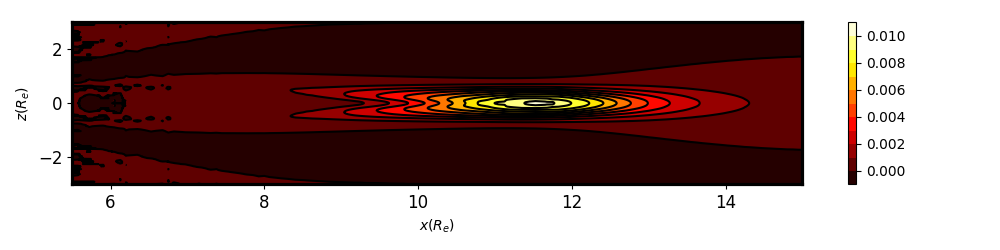}
  
  \includegraphics[width=\textwidth]{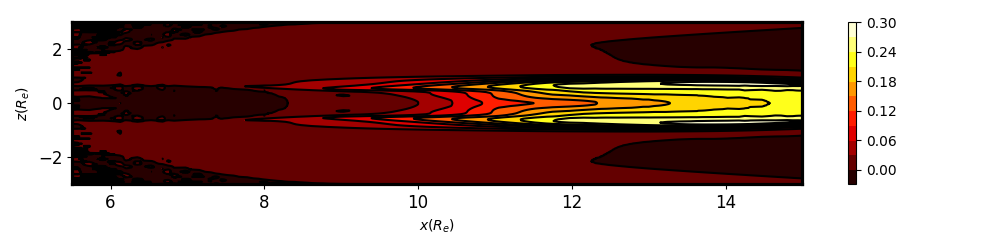}
  
  \includegraphics[width=\textwidth]{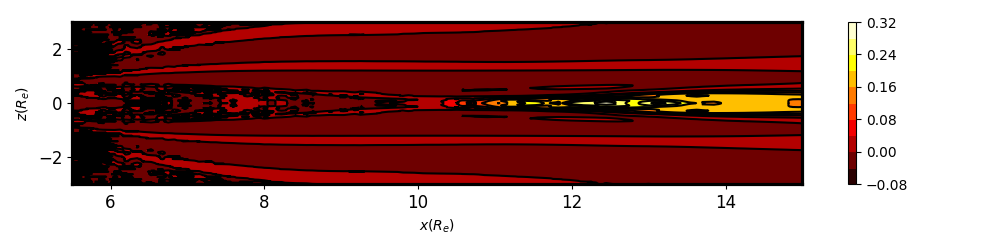}
\end{center}
\caption{Tailward flow contour in $x-z$ meridian plane ($y=0$) at times $t=400$ (upper), $t=470$ (middle), and $t=504$ (lower).}
\label{fig:sm_ux}
\end{figure}

Nonlinear development of ballooning instability leads to reconnection and plasmoid formation in near-Earth magnetotail finger structure and plasmoid structure (Fig.~\ref{fig:sm_pres_bline}). However, this does not visibly take place for a long period of linear growth phase. The magnetic configuration and pressure distribution remain nearly unperturbed well through the time $t=350$. In the $x-y$  equatorial plane ($z=0$) of magnetotail, the pressure contour evolution shows the development of sinusoidal variation, with the same initial periodicity in the $y$-direction in the early nonlinear growth phase around the time $t=400$ (Fig.~\ref{fig:sm_pres_bline}, 1st panel), which quickly transitions into the narrowed tailward protruding finger structure in the nonlinear saturation phase by the time $t=470$ (Fig.~\ref{fig:sm_pres_bline}, 3rd panel).  In the meanwhile, the initial closed field lines in the $x-z$ meridian plane ($y=0$) remain closed as the slow thinning of the entire current sheet progresses (Fig.~\ref{fig:sm_pres_bline}, 1st and 3rd panels). After the nonlinear perturbation passes its peak amplitude when $t\gtsim 470$, the narrowed tailward protruding finger structures in pressure contour retract to remnant variation in $y$-direction, whereas the continued thinning of current sheet around and tailward of $x=10$ evolves into the rapid emerging of multiple $X$-points and plasmoids in the late nonlinear phase about the time $t=504$ (Fig.~\ref{fig:sm_pres_bline}, 5th and 6th panels). Time evolution of tailward flow $u_x$ contour in $x-z$ meridian plane ($y=0$) shows that the associated tailward flow dominates the early ballooning development until the bipolar flow structure appears after the onset of reconnection (Fig.~\ref{fig:sm_ux}). 

\begin{figure}
\begin{center}
\includegraphics[width=\textwidth]{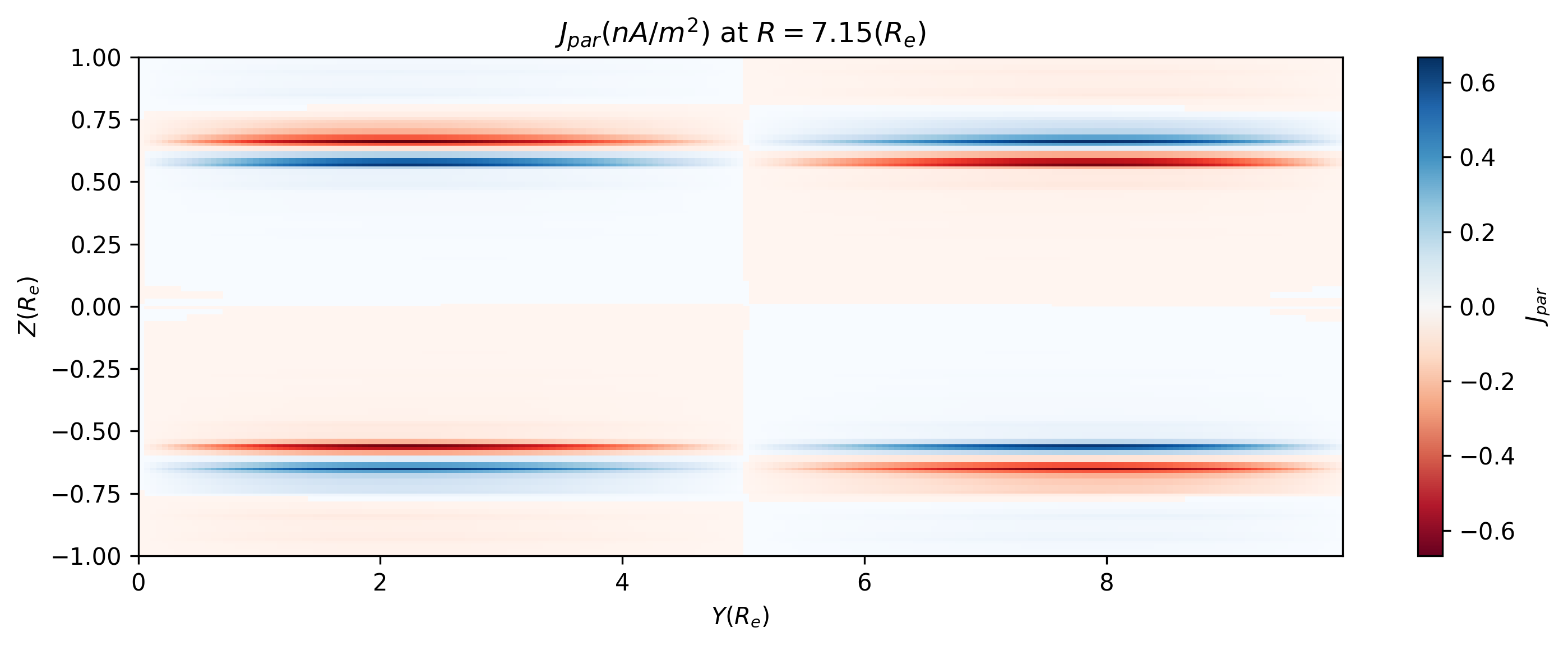}
    
\includegraphics[width=\textwidth]{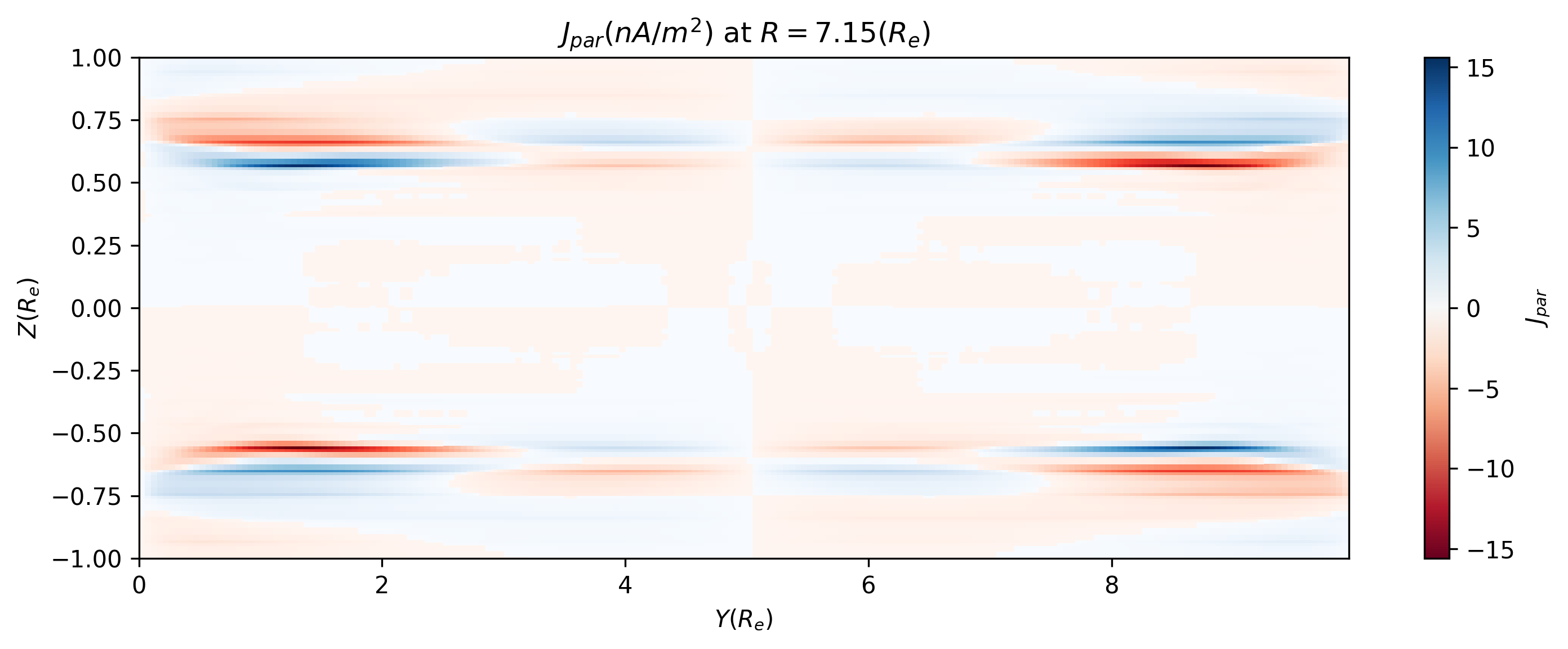}

\includegraphics[width=\textwidth]{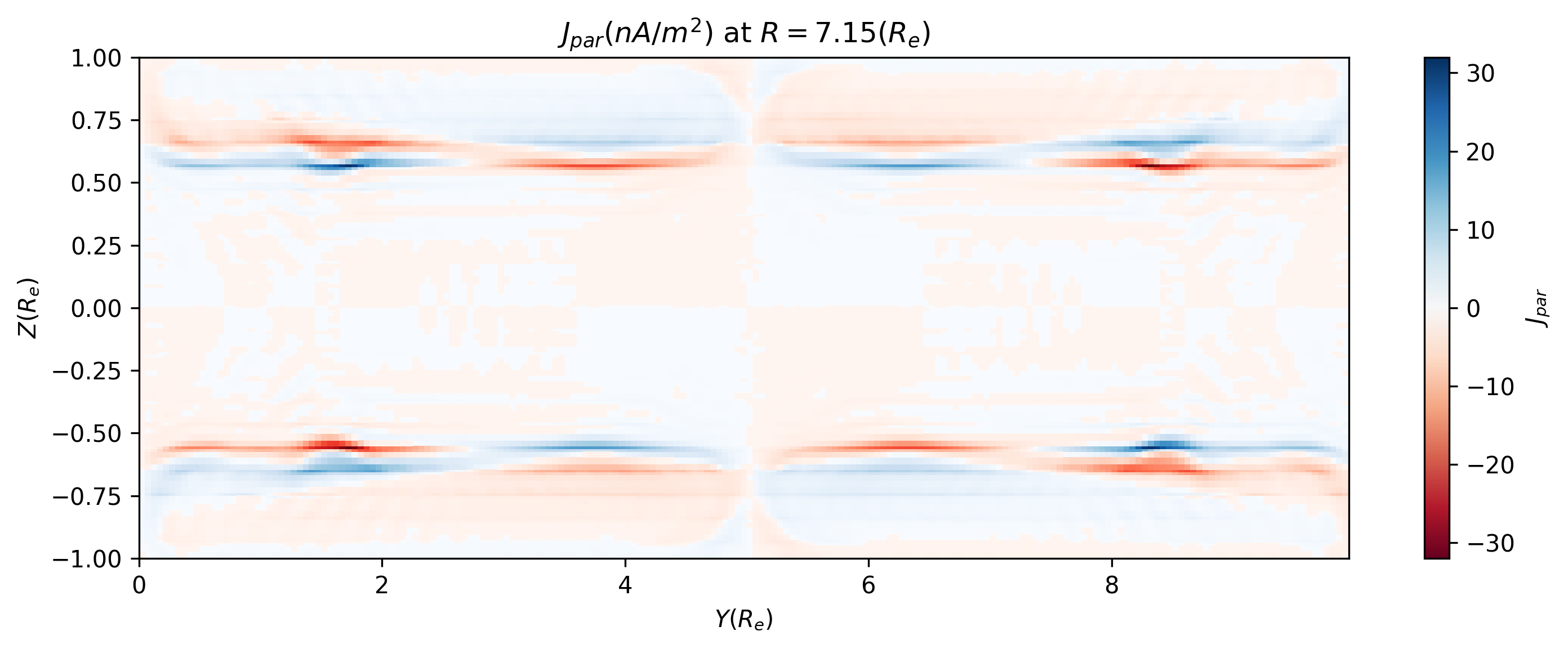}
\end{center}
\caption{FAC density distribution in $y-z$ plane ($x=7.15$) at $t=400$ (upper), $t=470$ (middle), and $t=504$ (lower).}
\label{fig:sm_fac}
\end{figure}

The field-aligned current (FAC) density at near-Earth side boundary plane (at $x=7.15$)  of the simulation domain for magnetotail has been evaluated based on the MHD simulation results and the $y-z$ plane contour of FAC density shows its periodic variation pattern in the dusk-dawn ($y$) direction within a narrow band around $z=\pm 0.65$ (Fig.~\ref{fig:sm_fac}). At the earlier time around $t=350$, the FAC periodicity in $y$-direction is monochromatic with the mode number $n=1$, which later evolves to a mixture of multiple harmonics with dominant $n=2$ component after $t=470$. It is worth noting that there is no significant north- or south-ward expansion of the FAC structure, which remains stationary in the narrow band throughout the development of ballooning instability. The mapping of such an FAC through its precipitation to the auroral zone in ionosphere is later used as input to TREx model to reconstruct the corresponding auroral structure for comparison with ASI observation images in Sec.~\ref{sec:mod_dat_com}. As discussed in more details there, the lack of  significant north- or south-ward expansion of the FAC structure or the corresponding pole-ward or Earth-ward expansion of auroral pattern motivates the consideration of an alternative scenario for ballooning instability development next.

\subsubsection{Double-mode initial perturbation}
\label{sec:double}
\begin{figure}
\begin{center}
\includegraphics[width=\textwidth]{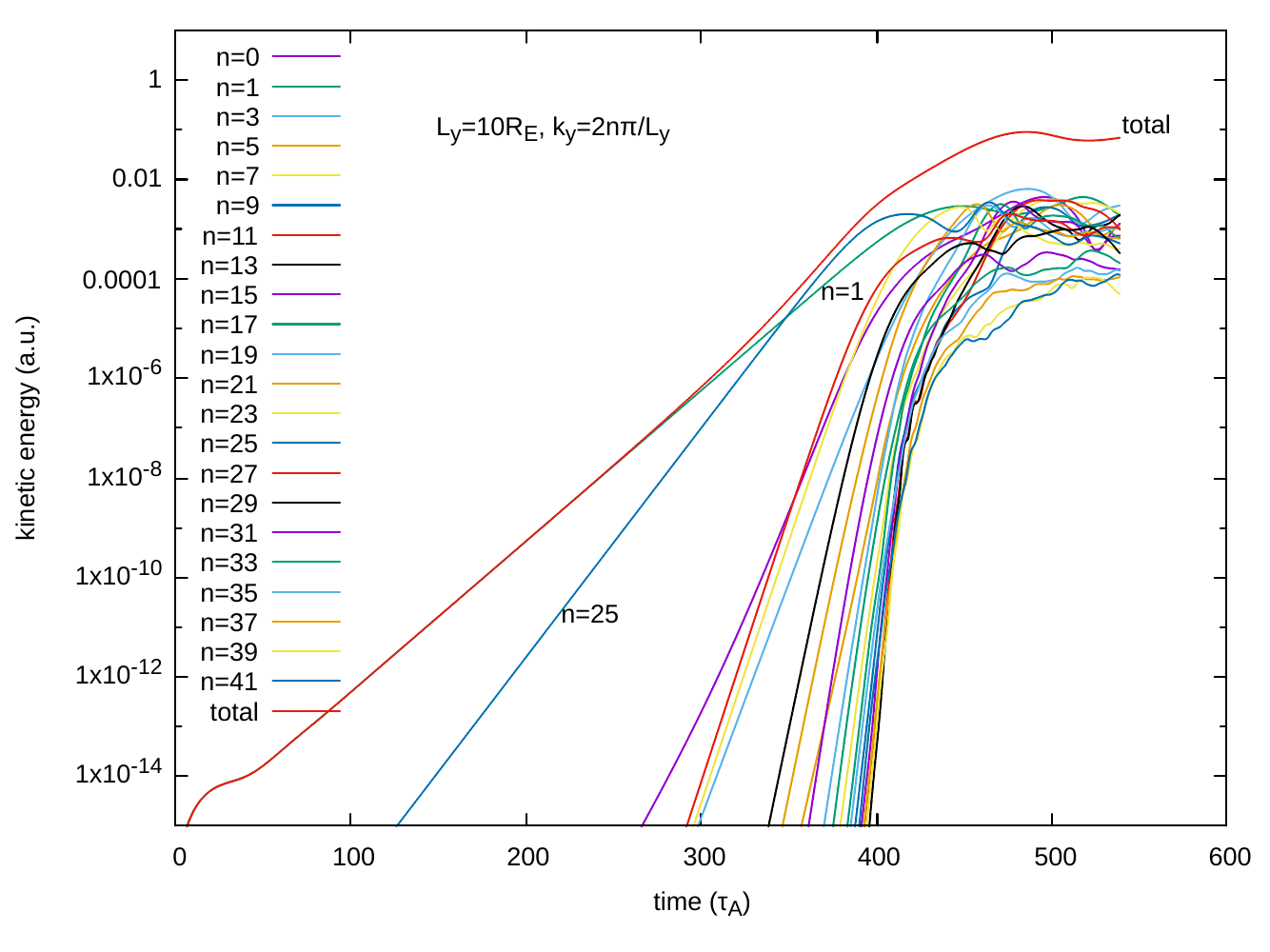}
\end{center}
\caption{Kinetic energies of all Fourier components of perturbation as well as their sum as functions of time, where only the $n=1$ and the $n=25$ components have non-zero amplitudes of  $4.5\times 10^{-9}$ and $2.5\times 10^{-11}$ respectively in the initial perturbation.}
\label{fig:dm_ken_growth}
\end{figure}

We now proceed with the discussion on another scenario where a non-monochromatic perturbation with two Fourier components initiates in the near-Earth magnetotail development. In particular, the initial magnetic perturbation includes a dominant mode with the mode number $n=1$ (i.e. wavelength $L_y=10R_e$) and the amplitude $4.5\times 10^{-9}$, and another mode with $n=25$ (i.e. wavelength $L_y=0.4R_e$) and a weaker amplitude $2.5\times 10^{-11}$. Other features of the initial perturbation are set up similarly to the single-mode case in the previous subsection with slight variations. For example, the perturbation also assumes a Gaussian profile along the $x$-axis that is peaked around $x=-10R_e$ but with a much wider half-height width $100R_e$. Whereas only two harmonics (i.e. $n=1$ and $n=25$) are initialized with nonzero amplitudes, the nonlinear simulation also includes and resolves a total of $43$ Fourier components with $n=0-42$ in the $y$-direction.  Both the dominant $n=1$ and the sub-dominant $n=25$ components grow exponentially initially, and due to the faster linear growth of $n=25$ component, its amplitude eventually surpasses that of the $n=1$ component after the time around $350\tau_A$. By the time both the $n=1$ and the $n=25$ components enter the middle of their exponential growth phases,  other Fourier components are sequentially excited and growth exponentially through nonlinear coupling and beating (Fig.~\ref{fig:dm_ken_growth}). These two components first reach saturation after about $400\tau_A$, whereas other components also enter saturation after about  $450\tau_A$, when the kinetic energies of  all perturbation components have become comparable within about one order of magnitude.

\begin{figure}
\begin{center}
\includegraphics[width=0.85\textwidth]{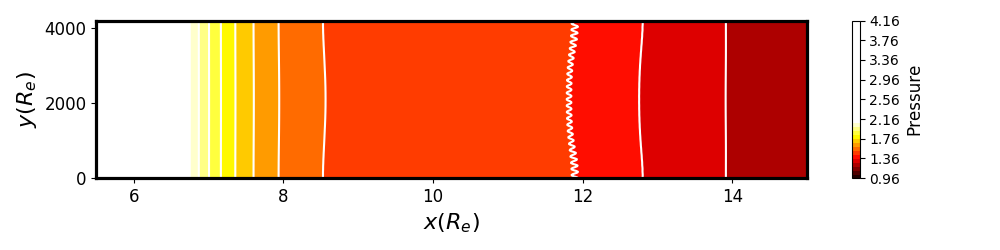}

\includegraphics[width=0.85\textwidth]{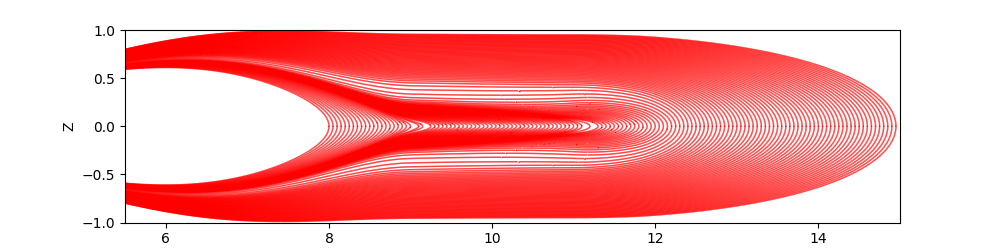}
  
\includegraphics[width=0.85\textwidth]{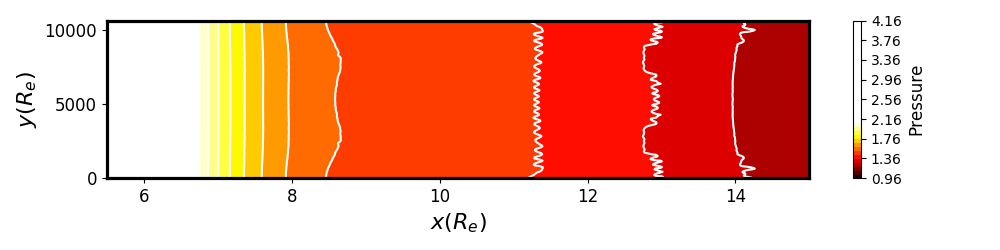}

\includegraphics[width=0.85\textwidth]{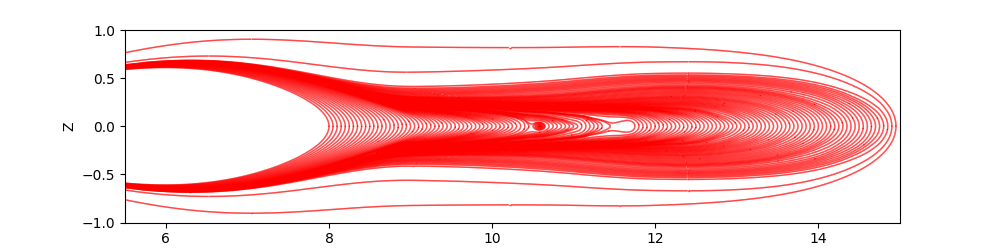}

\includegraphics[width=0.85\textwidth]{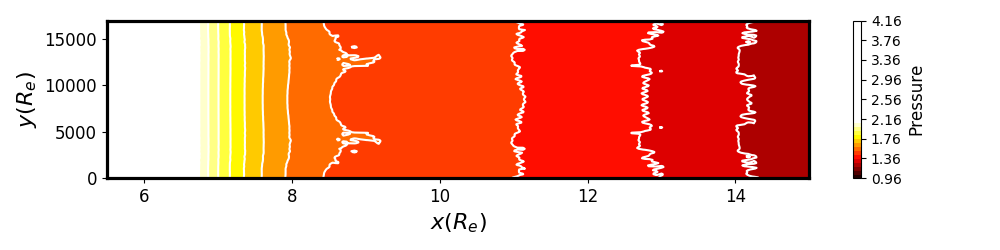}

\includegraphics[width=0.85\textwidth]{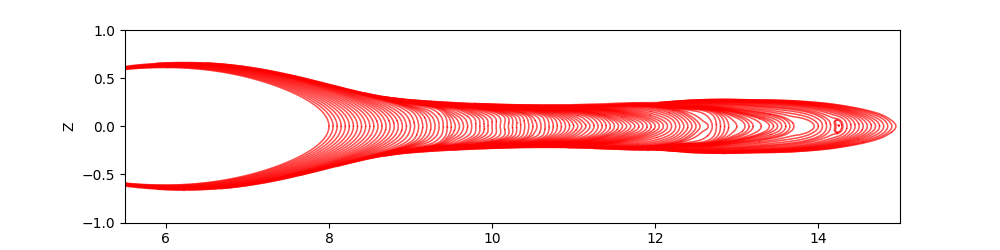}
\end{center}
\caption{Pressure contour in $x-y$ equatorial plane ($z=0$) (1st, 3rd, and 5th rows) and magnetic field lines in $x-z$ meridian plane ($y=0$) (2nd, 4th, and 6th rows) at times $t=400$ (1st and 2nd rows), $t=465$ (3rd and 4th rows), and $t=502$ (5th and 6th rows).}
\label{fig:dm_pres_bline}
\end{figure}

Due to the presence and faster growth of the $n=25$ component, the development of the $n=1$ ballooning finger pattern in the pressure perturbation contour within the equatorial plane ($z=0$) of magnetotail is soon superimposed by the shorter wavelength in the $y$-direction (i.e.~$0.4R_e$) from the $n=25$ component since the middle exponential growth phase after  the time $350\tau_A$  (Fig.~\ref{fig:dm_pres_bline}, 1st panel). In the $x-z$ meridian plane ($y=0$), the closed field lines in the near-Earth region around  $x=10$ quickly evolve into a much thinner current sheet and maintains its width throughout the exponential and the early saturated growth phase (Fig.~\ref{fig:dm_pres_bline}, 2nd panel). By the time of early saturation stage around $t=465$, the pressure contour develops non-sinusoidal structure at $x\gtsim 13$ and two plasmoids appear in the region between $x=10$ and $x=12$, whereas the pressure perturbation pattern remains sinusoidal in near-Earth region at $x\ltsim 12$ (Fig.~\ref{fig:dm_pres_bline}, 3rd and 4th panels). Well into the nonlinear saturation stage around $t=502$,  the pressure perturbation variation structures in $y-$ direction become non-sinusoidal in the entire near-Earth region from $x\gtsim 8$ to $x\gtsim 14$, and a plasmoid emerges around $x\gtsim 14$ further tailward of the earlier plasmoid locations at $t=465$ (Fig.~\ref{fig:dm_pres_bline}, 5th and 6th panels). In the meantime, the current sheet becomes thinner by this stage.

\begin{figure}
\begin{center}
 \includegraphics[width=\textwidth]{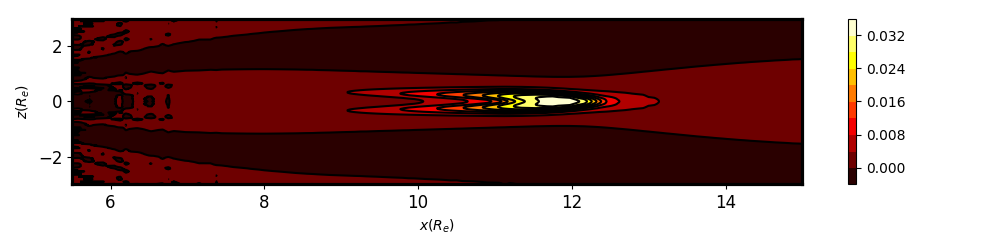}
  
 \includegraphics[width=\textwidth]{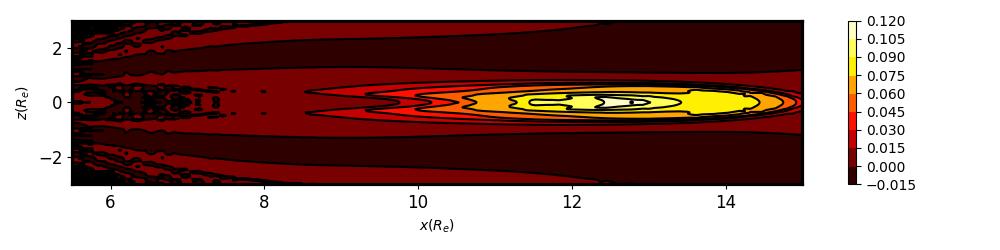}
 
 \includegraphics[width=\textwidth]{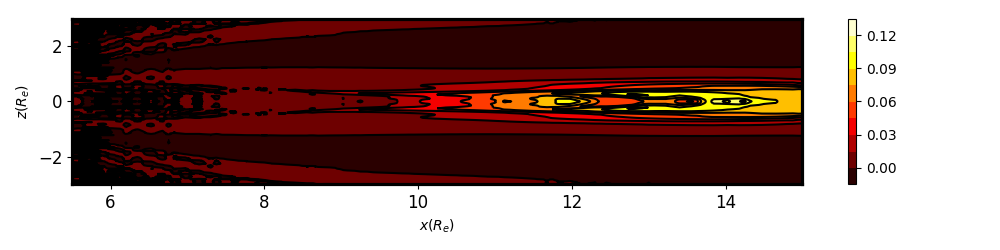}
\end{center}
\caption{Tailward flow contour in $x-z$ meridian plane ($y=0$) at times $t=400$ (upper), $t=465$ (middle), and $t=502$ (lower).}
\label{fig:dm_ux}
\end{figure}

Although the dominant flow near equatorial plane maintains the tailward direction during the process of above ballooning mode evolution, it is the structure of the relative flow within the reference frame of the average tailward flow that correlates well with the tailward ballooning front expansion and the plasmoid formation at the corresponding time moments (Fig.~\ref{fig:dm_ux}). For example, the disconnected bi-polar structures in the $u_x$ contour become apparent  around the locations of plasmoid formation at the time $t=465$ (Fig.~\ref{fig:dm_ux}, middle panel) and the time $t=502$ (Fig.~\ref{fig:dm_ux}, lower panel).

\begin{figure}
\begin{center}
\includegraphics[width=\textwidth]{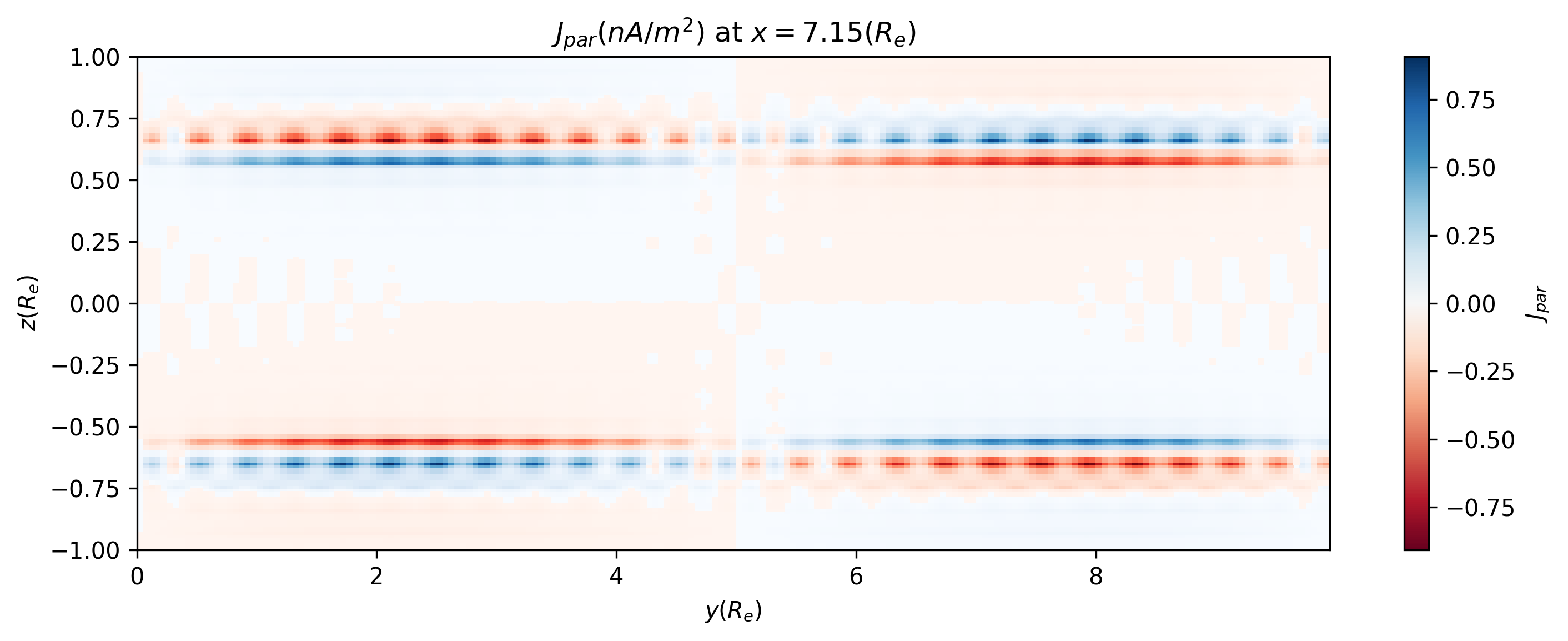}
    
\includegraphics[width=\textwidth]{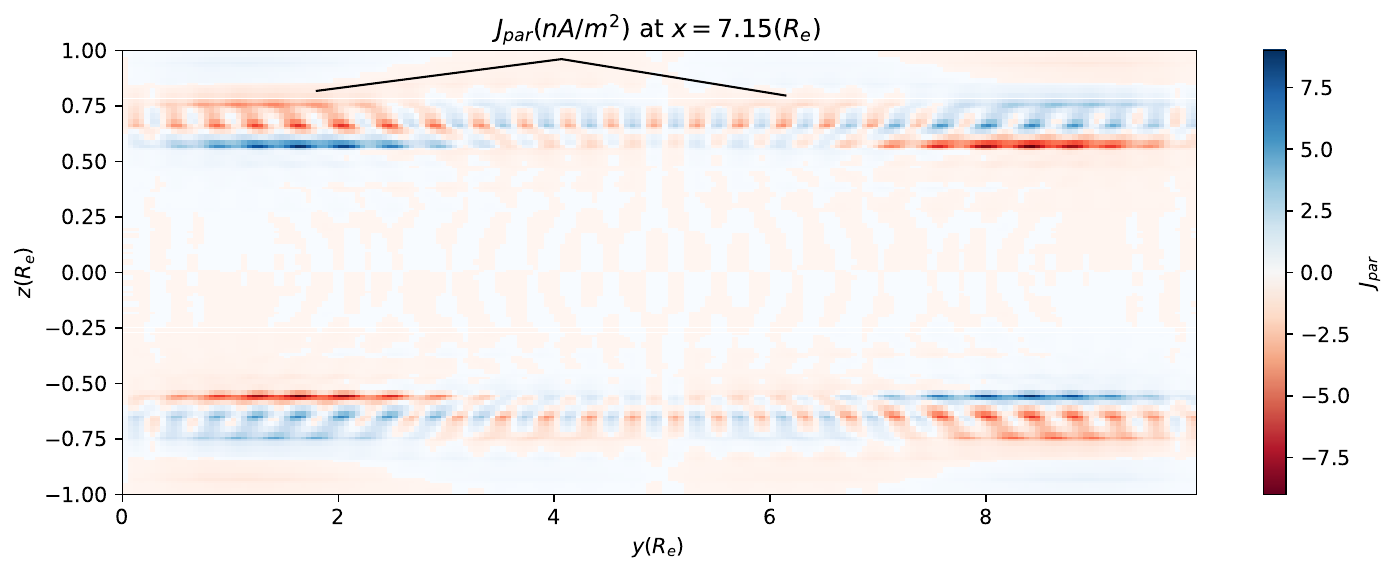}

\includegraphics[width=\textwidth]{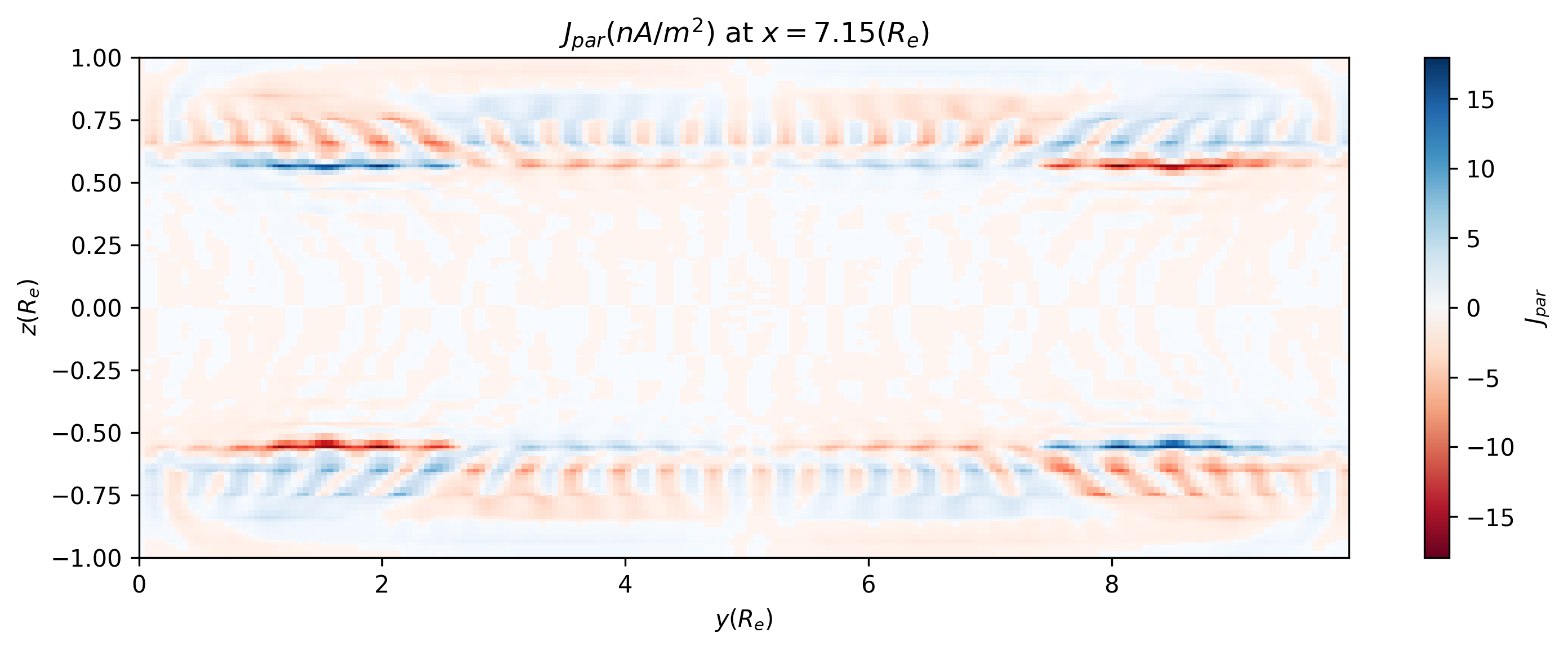}
\end{center}
\caption{FAC density distribution in $y-z$ plane ($x=7.15$) at $t=400$ (upper),  $t=465$ (middle), and $t=502$ (lower). In the middle panel ($t=465$), the two dark lines mark the FAC region that maps to the newly emerged poleward arc shown in the time frame of $1315.0$ seconds in Fig.~\ref{fig:dm_fac_map}.}
\label{fig:dm_fac}
\end{figure}

Following the same procedure, the FAC density at near-Earth side boundary plane (at $x=7.15$)  of the magnetotail simulation domain has been evaluated based on the MHD simulation results.  The periodic structures of FAC density in the dusk-dawn ($y$) direction are initially located within two narrow bands around $z=\pm 0.8$, and by the time $t=400$ the dominant wavelengths show those of both the $n=1$ and the $n=25$ components due to their comparable amplitudes at the beginning of the saturation phase (Fig.~\ref{fig:dm_fac} upper panel). At later time from $t=465$ to $t=502$, the FAC structure becomes dominated by the $n=25$ component with shorter wavelength in $y$-direction, and its band width in $z$-direction has apparently broadened to about $0.2R_e$ at each south and north symmetric location. Well into the saturation phase at $t\gtsim 502$, the FAC periodicity in $y$ also involves a mixture of multiple other harmonics.  In Sec.~\ref{sec:mod_dat_com}, we also use the mapping of such an FAC through its precipitation to the auroral zone in ionosphere as input to TREx model to reconstruct the corresponding auroral structure for comparison with ASI observation images, which indicates that the presence of the second harmonic with shorter wavelength in initial perturbation proves crucial for better agreement. 

\section{FAC auroral zone projections and comparison with observations}
\label{sec:iono}

In this section, we shall endeavor to compare the model simulation with the realistic THEMIS observations mainly from ground-based optical ASI, along with certain in-situ satellite data.

\begin{figure}
\begin{center}
\includegraphics[width=\textwidth]{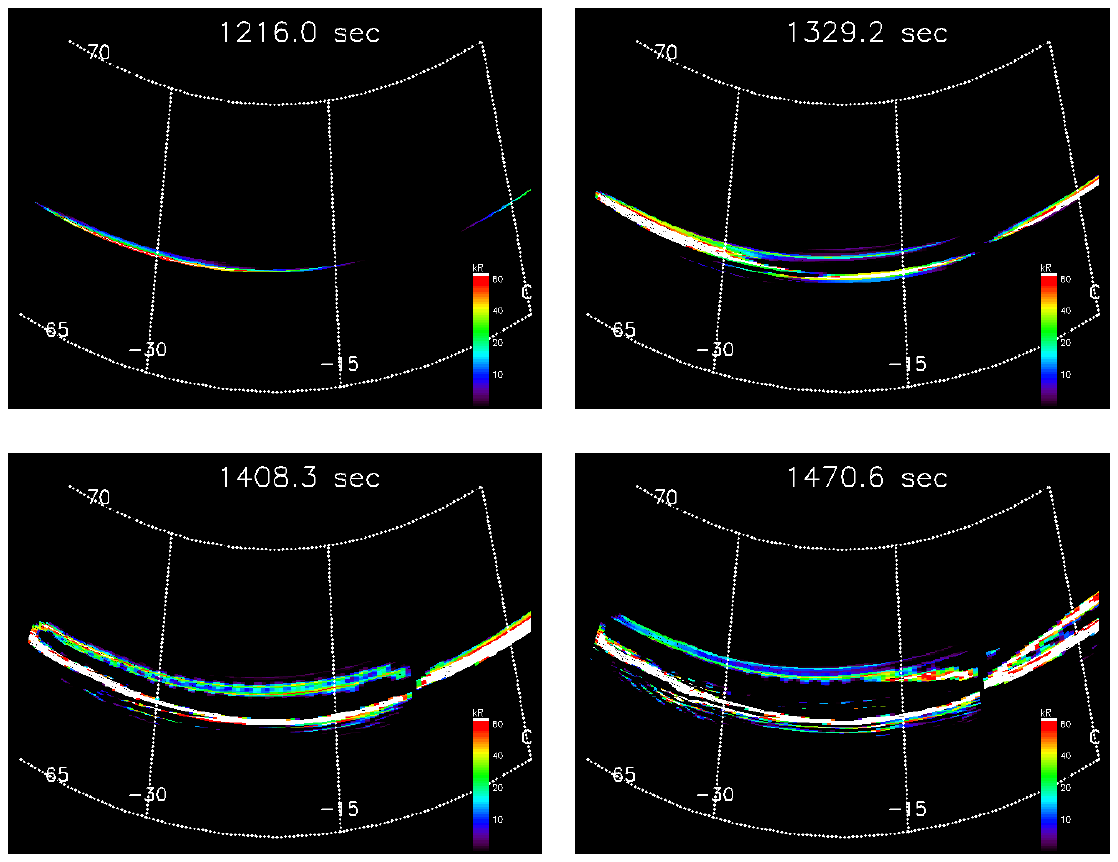}
\end{center}
\caption{Mapping of FAC/precipitation to the ionosphere using TREx model and FAC data from the single-mode simulation case at a sequence of four time frames.}
\label{fig:sm_fac_map}
\end{figure}

\begin{figure}
\begin{center}
\includegraphics[width=\textwidth]{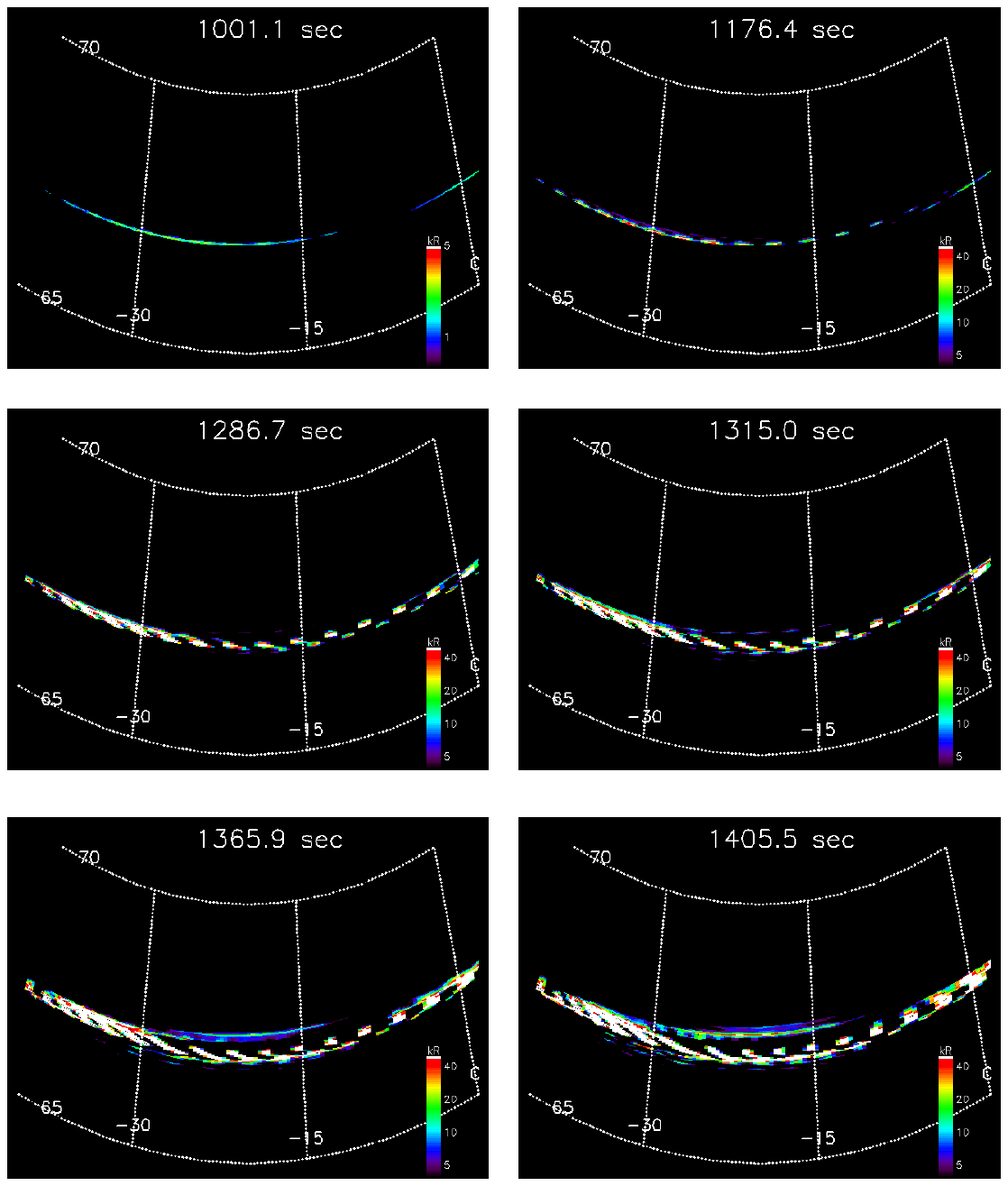}
\end{center}
\caption{Mapping of FAC/precipitation to the ionosphere using TREx model and FAC data from the double-mode simulation case at a sequence of six time frames.}
\label{fig:dm_fac_map}
\end{figure}

\begin{figure}
\begin{center}
\includegraphics[width=0.9\textwidth]{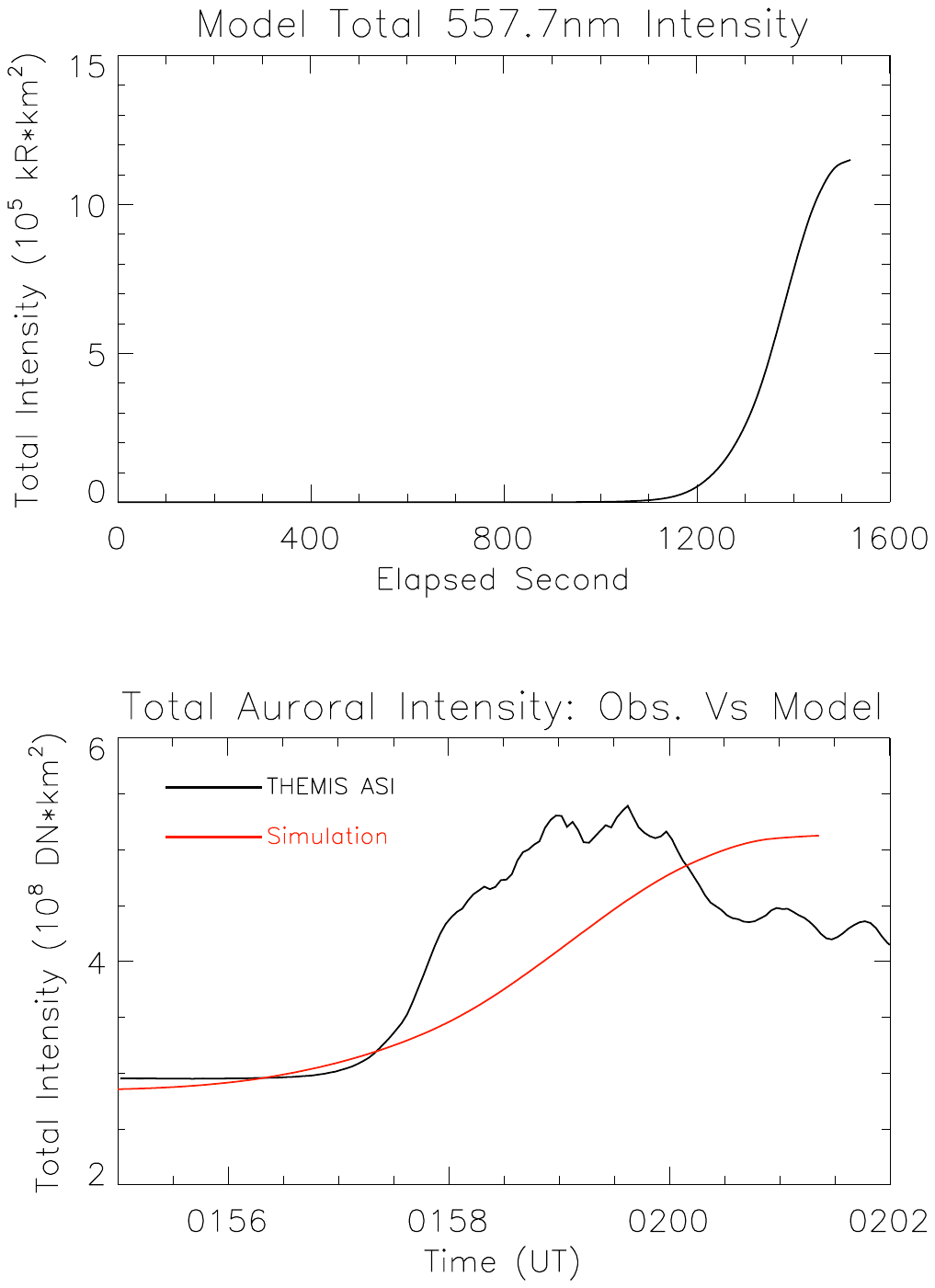}
\end{center}
\caption{The $557.7$ nm (upper) and the total auroral intensity (lower, red line) obtained from TREx model and FAC data from the double-mode simulation case as a function of time. The dark line in the lower panel is the total auroral intensity as a function of time from THEMIS ASI observation.}
\label{fig:trex_auroral_intensity}
\end{figure}

\subsection{FAC auroral zone projection}
\label{sec:fac_proj}
To compare our model output with optical auroral observations, we first need to perform a proper mapping between the magnetosphere and the ionosphere. As previously introduced, the current sheet model adopted in the simulation is a modified Harris sheet in Cartesian geometry (see Sec.~\ref{sec:initial}). While the current sheet model considers the increase of $B_z$ at inner distance, there is no magnetic field from the Earth, i.e. a quasi-dipolar component.   To achieve a more realistic mapping, we assume a gradual transition of the current sheet to a Tsyganenko 89 (T89) model~\citep{tsyganenko89a} at the inner edge of the current sheet. Note that the reference points ($x=0$ and $y=0$) in the definition of (\ref{eq:bzmin_pw}) are arbitrary but must be specified when performing a realistic mapping. In practice, we set $x=0$ and $y=0$ to physical GSM coordinates $X_{\rm GSM}=0.5 R_{\mathrm E}$ and $Y_{\rm GSM}=1 R_{\mathrm E}$, respectively. This is meant to achieve a better agreement between the current sheet $B_z$ component at $\sim 8 R_{\mathrm E}$ and the TH-A observation (see Fig.~\ref{fig:themis_orbit}),  along with the consideration that substorm onset is known to be biased towards the pre-midnight sector. To depict a transition from the Cartesian model current sheet to a quasi-dipolar field closer to the Earth, we assume the following form of vector potential,
\begin{equation}
\mathbf{A}=\alpha_1(x)\mathbf{A}_{\rm CS}+(1-\alpha_1)\mathbf{A}_{\rm T89}
\end{equation}
where $\alpha_1=\sin^2{[(X_{\rm GSM}+6)\pi/3.3]}$ is so specified that it varies from $1$ at $X_{\rm GSM}=-7.65 R_{\mathrm E}$ to $0$ at $X_{\rm GSM}=-6$, the model current sheet is overall shifted slightly to the dusk, so that $y=0$, the azimuthal center of the current sheet, corresponds to $Y_{\rm GSM}=1 R_{\rm E}$. $X_{\rm GSM}=-7.65 R_{\rm E}$  is the location we retrieve the FACs from the model simulation and map to the ionosphere. The formula and involved parameters required to construct the T89 vector potential $\mathbf{A}_{\rm T89}$ are from~\citet{tsyganenko89a}. $K_p=3$ is assumed. Since the model current sheet is symmetric in the $z$-direction, we have also assumed a zero dipole tilt in constructing $\mathbf{A}_{\rm T89}$.

The next step is to evaluate the electron precipitation parameters incident upon the ionosphere from the modeled FAC. An auroral acceleration process has to be considered for such a purpose. Auroral acceleration, especially the quasi-static acceleration, remains inadequately understood to date. In this study, we shall adopt the common approach by assuming a Knight relation between the magnetosphere FAC and the parallel potential. Mathematical details and equations for such a procedure are comprehensively described elsewhere (e.g.~\citep{fridman80a,raeder00a,zhangb15a}). To evaluate the parallel potential drop, the magnetosphere electron density and temperature are required. While the electron density can be given by the model,  the electron temperature cannot be directly obtained from a MHD model, and certain assumption has to be involved (e.g.~\citep{zhangb15a}). In this study, based on in-situ observations of TH-D, which maps closest to the breakup arc, we set a fitted electron temperature of $2.4 keV$ in the Knight relationship (see Fig.~\ref{fig:themis_orbit}d).

After obtaining the parallel potential, we assume a shifted Maxwell flux spectrum of electron precipitation and pass the spectrum to the TREx-ATM model~\citep{liang16a,liang17a,liang25a} to compute the resulting spatio-temporal variations of optical auroral intensities. We shall use the green-line (557.7 nm) intensities in the following presentation since: (a) the green-line auroras are typically dominant in optical brightness; (b) the 557.7 nm intensity is found to have a reasonable correlation with the THEMIS white-light data count~\citep{gabrielse21a}, facilitating a quantitative comparison.

\subsection{Model-data comparison}
\label{sec:mod_dat_com}
We first briefly check the simulated optical auroras in the single-mode run. In such a single-mode run, the lower-order (larger-wavelength) modes intensify first and dominate (see Fig.~\ref{fig:sm_ken_growth}), until late in the elapsed time. As one can see from Fig.~\ref{fig:sm_fac_map}a and~\ref{fig:sm_fac_map}b, a large-scale arc spanning over $>20^\circ$ MLON, appears first and grows to an intensity of several tens of $kR$. This is unrealistic for the growth phase arc.  Clues of azimuthal-spaced structures with wavelengths of $\sim 30^\circ$ MLON indeed show up later, hinting at the intensification of higher-order modes. However, these smaller-scale structures are $\sim 1^\circ$ MLAT poleward of the equatorward-most arc (Fig.~\ref{fig:sm_fac_map}c), and appear to decline in $\sim 1$ minute (Fig.~\ref{fig:sm_fac_map}d). The simulated auroral sequence of the single-mode ballooning instability thus does not resemble the realistic optical observations of the substorm onset.

We then look into the simulated auroras of the double-mode run. A full movie showing the time evolution of the simulated auroras is given in the supplementary material. Fig.~\ref{fig:dm_fac_map} exemplifies six selected frames, aiming to make a direct analogy to the sequence shown in Fig.~\ref{fig:asi_20090305}. A larger-scale arc ($>15^\circ$ MLON) with weaker intensity (a few $kR$, Fig.~\ref{fig:dm_fac_map}a) appears first, which may correspond to the later-growth-phase arc. Later, pronounced beading structures with stronger intensities (tens of $kR$, note the color scale difference between Fig.~\ref{fig:dm_fac_map}a and the rest of the panels) emerge at the same latitude as the preexisting arc. The beading structures continue to intensify in the next couple of elapsed minutes and, due to the ongoing growth of higher-order modes, feature smaller azimuthal distances between successive beads. As can be inferred from Fig.~\ref{fig:dm_fac_map}c,  in the $-2^\circ$ to $0^\circ$ MLON range, the azimuthal separation of the beads is estimated to be $\sim 1.5^\circ$ MLON. This is slightly larger than, yet fairly comparable to, that inferred from THEMIS ASI observations ($\sim 1^\circ-1.3^\circ$ MLON). We note that here the finest resolved scales of azimuthal "wavelength" in the simulated auroral structure, when mapped to the tail, tend to be smaller than the wavelength of the $n=25$ mode in the initial perturbation. This indicates that the dominant modes at play here are those higher harmonics stemming from the nonlinear evolution of ballooning instability (See also Fig.~\ref{fig:dm_ken_growth}). About $\sim 3$ minutes after the appearance of beading structures, a new “arc” starts to develop at $\sim 0.3^\circ$ MLAT poleward of the initial beading arc. This new arc is initially weak but quickly brightens in the following minute; it is not homogeneous but also contains fine-structures. The above features are in good agreement with those unveiled in THEMIS ASI observations, as shown in Fig.~\ref{fig:asi_20090305}. The emergence timing of the new beading arc coincides with that of the formation of plasmoid in the magnetotail during nonlinear ballooning evolution.

We shall then compare the absolute intensity and the growth time of the auroral breakup. The total auroral intensity during a substorm breakup is an established way to evaluate the growth time of auroras and the underlying magnetosphere instabilities (e.g.~\citep{liang08a,liang17a,nishimuray16b,kalmoni17a}). To evaluate the total auroral intensity and the growth timescale of the substorm auroras from THEMIS ASI observations, we sample the auroras in a box shown as a dotted rectangle in Fig.~\ref{fig:asi_20090305}, at the 3-sec cadence of ASI. The sample region is chosen to cover a main portion of the breakup auroras yet not much affected by the moonlight contamination. The time evolution of the total modeled 557.7 nm intensity integrated over the simulation region is shown in Fig.~\ref{fig:trex_auroral_intensity}a. To compare this absolute intensity with the THEMIS ASI observation, we take the following steps. First, our simulation region is larger than the practical area to sample the ASI observations (see the dotted box in Fig.~\ref{fig:asi_20090305}); we thus proportionally scale the total auroral intensity in our simulation to the actual sample area. The key step in such a comparison effort is to establish a quantitative relation between the green-line auroral intensity and the THEMIS ASI data count (DC). This was done in~\citep{gabrielse21a} based on a fitting between THEMSI ASI DC and the absolute auroral intensities observed by meridian scanning photometers,
\begin{equation}
{\rm DC}={\rm DC}_{\rm BKG}+\left(\frac{I_{5577}}{15.85}\right)^{1/0.93}
\label{eq:trex_dc}
\end{equation}
in which ${\rm DC}_{\rm BKG}$ denotes the “background” DC consisting of dark currents, ambient non-auroral lights (e.g., moonlit), and the diffuse auroral intensity that is not accounted for in our simulation. In our data processing,  ${\rm DC}_{\rm BKG}$ is estimated from the average ASI DC in the upper portion of the sample box ($\sim 67^\circ$ MLAT, poleward of the preexisting arc) over $2$ minutes prior to the onset time.

Following the above procedures, we convert the total $557.7 nm$ intensity obtained from our model simulation to the ASI DC using Eq.~(\ref{eq:trex_dc}). In Fig.~\ref{fig:trex_auroral_intensity}b, we plot together the raw integrated DC sampled from THEMIS ASI, and that converted from our simulation. As one can see, the simulation reproduces the peak auroral intensity level fairly well, yet the growth timescale is slower by a factor of $\sim 2$. We conclude that the double-mode run reproduces, to a reasonable degree, the general pattern and overall intensity of the observed breakup auroras in the first few minutes. Conceptually, in the presence of auroral acceleration, the total precipitation energy flux, in turn, the auroral intensity, is supposed to be roughly proportional to the square of the upward FAC intensity (e.g.~\citep{fridman80a}). Since the FAC intensity obtained from simulation is a measure of the nonlinear magnetic perturbation level, which ultimately depends on the amplitude of initial perturbation, the overall agreement between the double-mode run and the observation in terms of the peak auroral intensity level justifies the setup of initial perturbation amplitude in simulations.

\begin{figure}
\begin{center}
\includegraphics[width=1.1\textwidth]{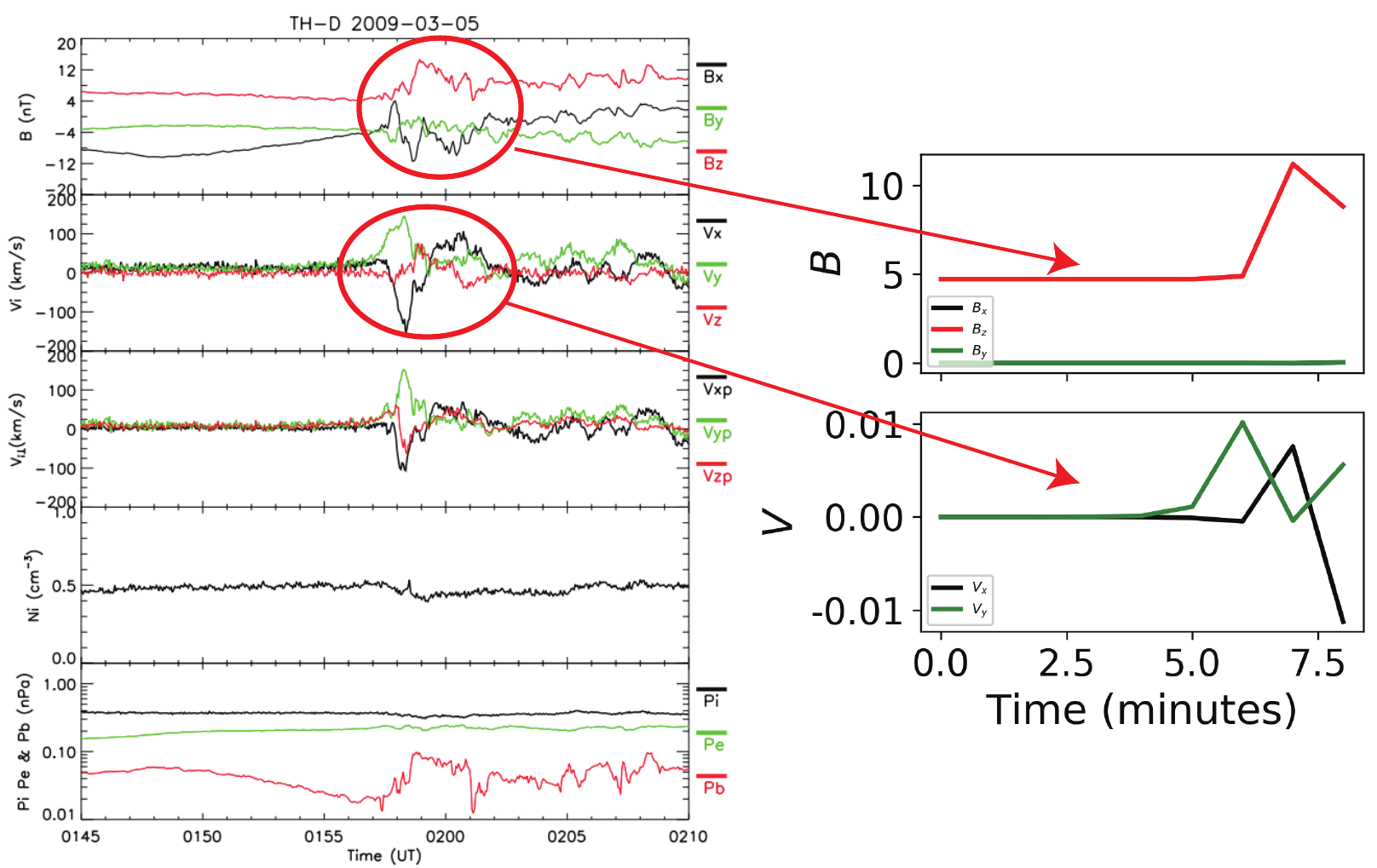}
\end{center}
\caption{Comparison between the TH-D observation data (left) and the NIMROD simulation results (right) on the in-situ magnetic field  and flow velocity along the TH-D orbit during the 20090305 substorm event from 0155UT to 0202UT.}
\label{fig:th_d_nim_20090305}
\end{figure}

In passing, it is worth noting that the comparison between the in-situ THEMIS satellite (e.g. TH-D) observation and the simulation data for the double-mode case also shows consistency and agreement during the time interval from 0155 UT to 0202 UT~(Fig.~\ref{fig:th_d_nim_20090305}). In particular, the relative magnitudes and time derivatives of three magnetic components at the time of dipolarization indicated by the sudden growth of $B_z$ component are well reproduced in simulation. Similarly, the simulation has also captured the sudden enhancement of Earthward flow at the time of dipolarization and the subsequent opposite switching of Earthward flow phase observed by TH-D. These in-situ comparisons provide additional support to the double-mode scenario that is mainly favored by the auroral observation-model comparisons. Not all components of magnetic field and flow velocity agree well, which is reflective of the over-simplifications in the MHD model and the generalized Harris sheet model adopted for simulations. On the other hand, the overall comparison also suggests that the $B_z$ and the $u_x$ components may be more relevant in determining the FAC and the auroral intensity structure.

\section{Summary and discussion}
\label{sec:sum}
In summary,  in this work we have reconstructed and identified the auroral signatures for the ballooning instability and plasmoid formation processes in the near-Earth magnetotail using a combined simulation method based on the resistive MHD model in NIMROD code and the auroral transport model TREx-ATM. Based on a set of THEMIS substorm onset events with good conjunction of auroral observations, our resistive MHD simulations have obtained the full temporal-spatial development of ballooning instability and plasmoid formation processes in the near-Earth magnetotail. The FAC density evaluated at the Earth side boundary of the magnetotail domain of simulation is used to compute the auroral emission density using the TREx-ATM code. The computed auroral evolution has revealed not only the beading structure but also the poleward expansion of the beading band and the later appearance of disconnected auroral segments. These beading and segmented auroral structures are found to be the direct mapping consequence of the FACs induced by the ballooning and plasmoid structures in the magnetotail respectively.

While the study performed in this work reinforces the notion that ballooning instability is a promising candidate for explaining the auroral beads at substorm onset, two new aspects are unveiled from our simulation for the first time. First, we find that the single-mode ($\lambda=10 R_{\rm E}$) ballooning instability, while also predicting the emergence of azimuthal quasi-periodic structures,  is unable to reproduce the correct evolution sequence of the beads and the preexisting arc.  To produce a reasonable evolution, a higher-order ($\lambda=0.4 R_{\rm E}$ assumed in this study) initial perturbation is required for the system. Such a shorter-wavelength perturbation is presumably externally driven. It is beyond the scope of this paper to explore in detail the potential driver of these shorter-wavelength perturbations in the tail, but we shall mention two possible candidates, (a) the ULF waves that were known to exist in the late growth phase~\citep{uritsky09a,motoba15a}; and (b) the BBFs and/or their azimuthal distributed “wedgelets” or “bundles” (e.g.~\citep{liuj13a,liuj15a}.

Second, in this study, we are not limited to initiation of ballooning instability and the onset auroral beads, but have further investigated the subsequent development of ballooning instability and its potential manifestation in the dynamics of substorm auroral expansion. In particular, we have found that in the double-mode initiation scenario, plasmoids start to form at NENLs around $10-12 R_{\rm E}$ at the beginning of the ballooning saturation phase around $t=460$. Near the earthward side of such NENLs, a thin FAC sheet is formed and leads to the emergence of a weaker, thin arc whose latitude is close to, yet somehow higher than, that of the initial beads. This is corroborated in the THEMIS ASI observation. The above observation and simulation align well with the result and interpretation in \citep{nishimuray25a}. In particular, the initial NENL is evaluated to be around $10-12 R_{\rm E}$ in our simulation (see Figs.~\ref{fig:dm_pres_bline} and~\ref{fig:dm_ux}), in good agreement with that ($\sim 11.8 R_{\rm E}$) estimated in \citep{nishimuray25a} from THEMIS observations.

We however note that, the bead wavelength and growth rate obtained in the current simulation are not exactly the same as those in observations. We speculate that such inconsistency might be partly due to the limitations of the MHD model used in this work. So far the resistive MHD simulations in this work are based on the single-fluid model where the two-fluid and finite-Larmor radius (FLR) effects are ignored.  The ion Larmor radius in the near-Earth magnetotail region is in the order of $0.1R_e$, which is around the minimum spatial scale resolved in the MHD simulation in this work.  Since the two-fluid and FLR effects naturally determine the cut-off wavelength for ballooning instability growth and provide the necessary mechanism for the fast reconnection rate, it is our next step to use the two-fluid MHD model with gyro-viscosity in order to include the lowest order and dominant two-fluid and FLR effects in simulations with finer spatial resolution up to scales of sub-ion gyro-radius in the magnetotail.  This would allow the identification of the finer scales of auroral structures that are of magnetotail origins during the substorm onset process.

\section{Data availability statement}
The THEMIS satellite and ASI data are available at themis.ssl.berkeley.edu and data-portal.phys.ucalgary.ca respectively. The simulation data are available at git.fusim.cn.

\section{Conflict of interest (COI) disclosure}
There is no potential conflict of interest (COI) that might affect the results reported in this manuscript or the development of this research. 

\begin{acknowledgments}
We are grateful for the support from the NIMROD team.  This research was supported by the National Natural Science Foundation of China Grant No. 41474143, the U.S. Department of Energy Grant No. DE-FG02-86ER53218, and the Hubei International Science and Technology Cooperation Project under Grant No.2022EHB003. Dr. Jun Liang and TREx-ATM are supported by the Canadian Space Agency. This research used the resources of the National Energy Research Scientific Center, a DOE office of Science User Facility supported by the Office of Science of the U.S. Department of Energy under Contract No. DE-AC02-05CH11231 using NERSC awards FES-ERCAP0024185, FES-ERCAP0027638, FES-ERCAP0031918, and FES-ERCAP0036510. The computing work in this paper is also supported by the Public Service Platform of High Performance Computing by Network and Computing Center of HUST.

\end{acknowledgments}

\end{article}
\end{document}